\documentclass[12pt,a4paper,dvips]{article}
\usepackage{a4p}
\usepackage{cite,mcite}
\usepackage{graphicx,rotating}
\usepackage{physics }
\usepackage{epsfig }
\usepackage{l3_titlePH,ifthen}
\usepackage{Lep}

\journalname{Phys. Lett. B}
      
\preprint{2005-012}       
\date{March 13, 2005}      

\Lep{2}

\def\Journal#1#2#3#4{{#1} {\bf #2} (#4) #3}

\def\NIMA{{Nucl. Instr. Meth.} A}

\def\PLB{{Phys. Lett.}  B}
\def\PRL{Phys. Rev. Lett.}
\def\PRD{{Phys. Rev.} D}
\def\ZPC{{Z. Phys.} C}
\def\PRP{{Phys. Rep.} C}
\def\CPC{Comput. Phys. Commun.}

\newlength{\capwidth}
\def\pb{\mbox{pb$^{-1}$}}
\newcommand{\EE}{\mathrm{e}^+\mathrm{e}^-}

\newcommand{\gamgam}{\gamma \gamma^*}

\newcommand{\roc}{\mathrm{\rho^\pm}}
\newcommand{\rocroc}{\mathrm{\rho^+\rho^-}}
\newcommand{\ro}{\mathrm{\rho^0}}
\newcommand{\roro}{\mathrm{\rho^0\rho^0}}
\newcommand{\pipi}{\mathrm{\pi^+\pi^-}}
\newcommand{\piz}{\mathrm{\pi^0}}
\newcommand{\pizpiz}{\mathrm{\pi^0\pi^0}}
\newcommand{\q}{ Q^2 }
\newcommand{\mgg}{ W_{\gamma \gamma }}
\newcommand{\ptt}{ p_t^2 }

\begin{document}

\begin{titlepage}
\title{Measurement of Exclusive  {\boldmath$\rho^+\rho^-$} Production \\
in Mid-Virtuality Two-Photon Interactions  \\
and Study of the {\boldmath$\gamgam \to  \rho\rho $} Process at LEP}

\author{The L3 Collaboration}

\begin{abstract}
Exclusive $\rocroc$ production in two-photon collisions between a
quasi-real photon, $\gamma$, and a mid-virtuality photon, $\gamma^*$, 
is studied with data collected
at LEP at centre-of-mass energies \mbox{$183 \GeV \le \sqrt{s} \le 209
\GeV{}$} with a total integrated luminosity of $684.8~\pb$.  The cross
section of the $\, \gamma \gamma^* \rightarrow \rocroc \,$ process is
determined as a function of the photon virtuality, $\q$, and the
two-photon centre-of-mass energy, $\mgg$, in the kinematic region:
$0.2 \GeV^2 \le \q \le 0.85 \GeV^2$ and $1.1 \GeV \le W_{\gamma\gamma} \le 3
\GeV$.
These results, together with previous  L3 measurements of $\roro$ and  
$\rocroc$ production,
allow a study of the $\gamgam \to  \rho\rho $ process
over the $\q$-region $0.2 \GeV^2  \le  \q  \le 30\GeV^2$.

\end{abstract}

\submitted

\vfill

\end{titlepage}

\section {Introduction}

The L3 Collaboration has recently measured
 the exclusive production of  $\roro$~\cite{L3paper269,L3paper292} 
  and $\rocroc$~\cite{L3paper287} pairs in the 
two-photon fusion process:
\begin{eqnarray}
\label{eq:eqn00}
\EE \to \EE \gamgam \to \EE \rho\rho \,,
\end{eqnarray}
\noindent
where the beam
electrons\footnote{Throughout this Letter, the term ``electron'' denotes both electron
 and positron.}  
radiate virtual photons which interact and produce a hadronic final state.
One of the photons, $\gamma$, is quasi-real,
characterised by a small value of its squared four momentum,
 $P_\gamma^2 = m_\gamma^2 \approx 0$, whereas the other one,
$\gamma^*$, has a significant virtuality,
$\q = -{P_{\gamma^*}}^2 =  -{m_{\gamma^*}}^2 \gg  -m_\gamma^2$.
Our measurements cover the two-photon centre-of-mass energy
\begin{equation}
\label{eq:rangewgg}
1.1 \GeV  \le   \mgg \le  3 \GeV .
\end{equation}
 
The two measurements  \cite{L3paper269, L3paper287}
done at large virtualities, $1.2 \GeV^2 \le \q \le 30 \GeV^2$,
provide a testing ground for a recently-developed 
QCD-based model~\cite{QCD}. 
This model describes well the $\q$-dependence of the $\roro$  production
at large momentum transfer~\cite{ANIKIN}. 
The measured cross sections  for $\roro$ and $\rocroc$ production
were found to have a similar dependence on $\mgg$ and to be of similar magnitude.
However,  the $\rocroc$ cross section is systematically higher than the $\ro\ro$ one.  
This is in contrast with the suppression and different $\mgg$
dependence of  $\rocroc$ production \cite{ARGUS} with respect to
 $\roro$ \cite{TASSO} observed in data with $\q \approx 0$ and $\mgg \le  2 \GeV$. 
We note that despite the wide range of theoretical models~\cite{ATTEMPTS,rosner}, 
$\rho$-pair production at  $\q \approx 0$ is still not well understood.
Therefore the experimental study of the $\q$-evolution of $\rho$-pair production is 
important to understand vector meson pair-production in two-photon interactions.

Previously, we performed a measurement of $\roro$ production~\cite{L3paper292}
 for intermediate  virtualities:
\begin{equation}
\label{eq:rangeq2}
0.2 \GeV^2  \le  \q  \le 0.85\GeV^2 . 
\end{equation}
\noindent 
In this Letter, we complement that study with the first measurement of the process:
\begin{eqnarray}
\label{eq:eqn01}
\EE \to \EE \gamgam \to \EE \rocroc
\end{eqnarray}
\noindent
in the kinematic region (\ref{eq:rangewgg}) and (\ref{eq:rangeq2}). 
These data  allow to follow the $\q$-evolution of the  $\rho\rho$-production
over two orders of magnitude in this variable.

The analysis techniques employed in this study are 
similar to those of our previous measurements \cite{ L3paper287, L3paper292}. 
The data used, corresponding  to an integrated luminosity of $684.8~\pb$,
 are the same as in Reference~\citen{L3paper292} and were collected 
by the L3 detector\cite{L3} at LEP
at centre-of-mass energies $183 \GeV \le \sqrt{s} \le 209 \GeV$.  
Scattered beam electrons
which have radiated photons
with virtualities in the range (\ref{eq:rangeq2}) can be  
``tagged''  by the Very Small Angle Tagger (VSAT) \cite{thesistasja}.
The VSAT is an electromagnetic calorimeter, constructed with  BGO crystals, 
with a geometrical acceptance covering the polar angle range 
 5 ${\rm mrad} \le \theta \le 10$ mrad, for
azimuthal angles in the ranges~ $-$1.25 rad$\,\le \!  \phi\! \le \,$1.25 rad
and \mbox{$\pi-$1.25 rad$\,\le \! \phi\!  \le \! \pi$+1.25 rad}.
When the electron with the largest scattering angle is detected in the
VSAT, the virtuality  of the photon it radiated  is, 
 within 1\% precision,  equal to the transverse momentum squared, $\ptt $, of
the final state hadron system:
\begin{equation}
\label{eq:eqn02}
\q = 2 E_b E_s (1 - \cos \theta_s) \approx E_b E_s \theta_s^2 \approx  \ptt  \,,
\end{equation}
where $ E_b$ is the beam energy, and $ E_s $ and $\theta_s $ are the
energy and the scattering angle of the tagged electron. Therefore
the VSAT is  not used to directly measure $\q $, but rather 
 to select exclusive final states by correlating the direction
of the transverse momentum vector of the tagged electron with the 
detected hadron system.

\section {Event Selection}

The reaction $\EE \to  \EE \rocroc$ contributing to the process
\begin{equation}
\label{eq:eqn06}
\EE \to \mathrm{e^+ e^-_{tag}}\pipi\pizpiz \,
\end{equation}
is identified by one and only one  scattered electron, $ \mathrm{e_{tag}}$, detected in the VSAT, 
two charged pions measured in the tracking chamber,
and energy clusters from the two-photon decays of the $\pi^0 $'s,
deposited in the BGO electromagnetic calorimeter.
These events are collected by two independent track-triggers
\cite{L3T}.  The trigger efficiency, as determined from the data
itself, is $(60  \pm 3 )\% $.

\begin{figure} [p]
  \begin{center}
\vskip -1cm
     \mbox{\epsfig{file=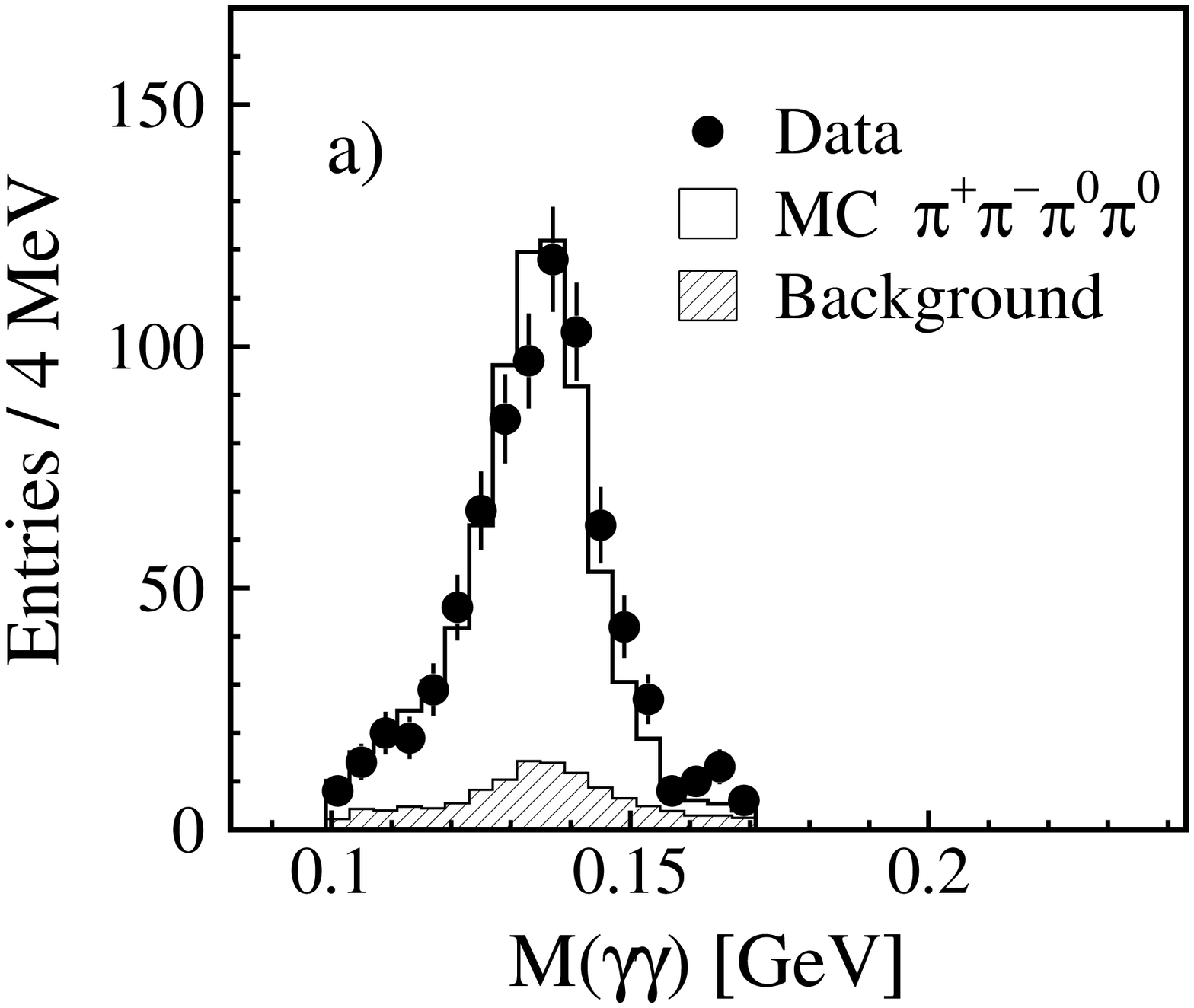,width=0.49\textwidth}}
     \mbox{\epsfig{file=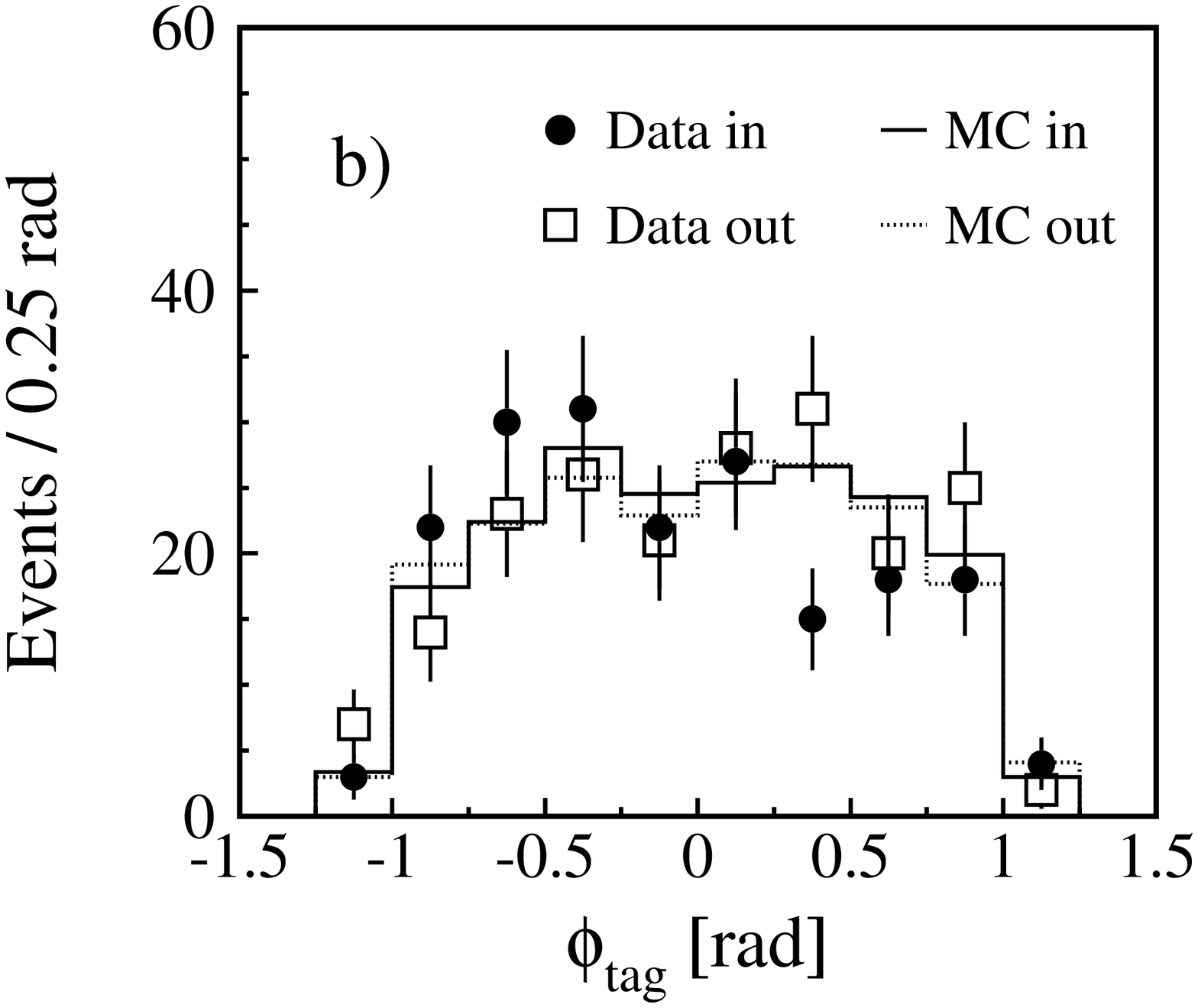,width=0.49\textwidth}}
\vskip -0.5cm
     \mbox{\epsfig{file=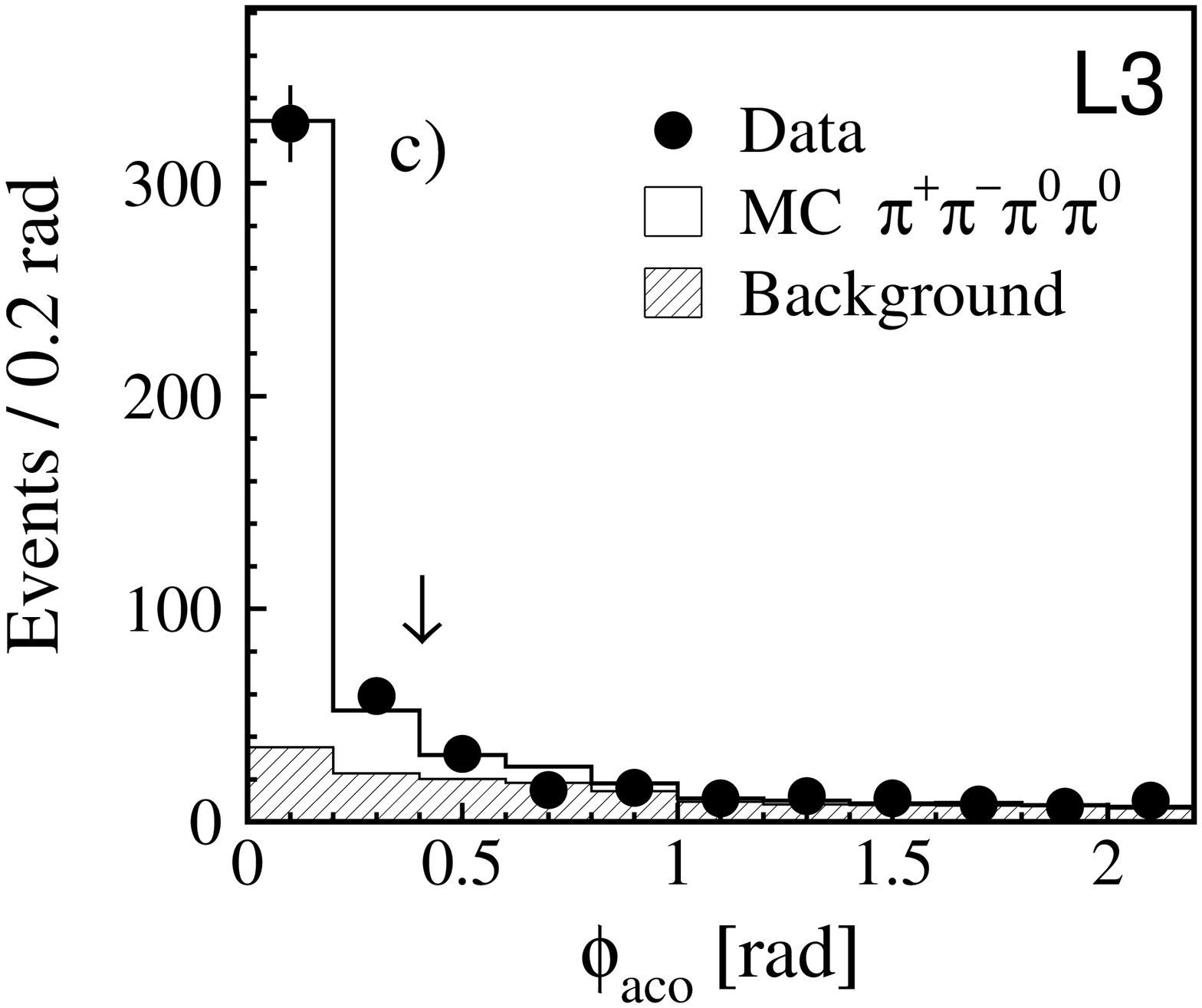,width=0.49\textwidth}}
     \mbox{\epsfig{file=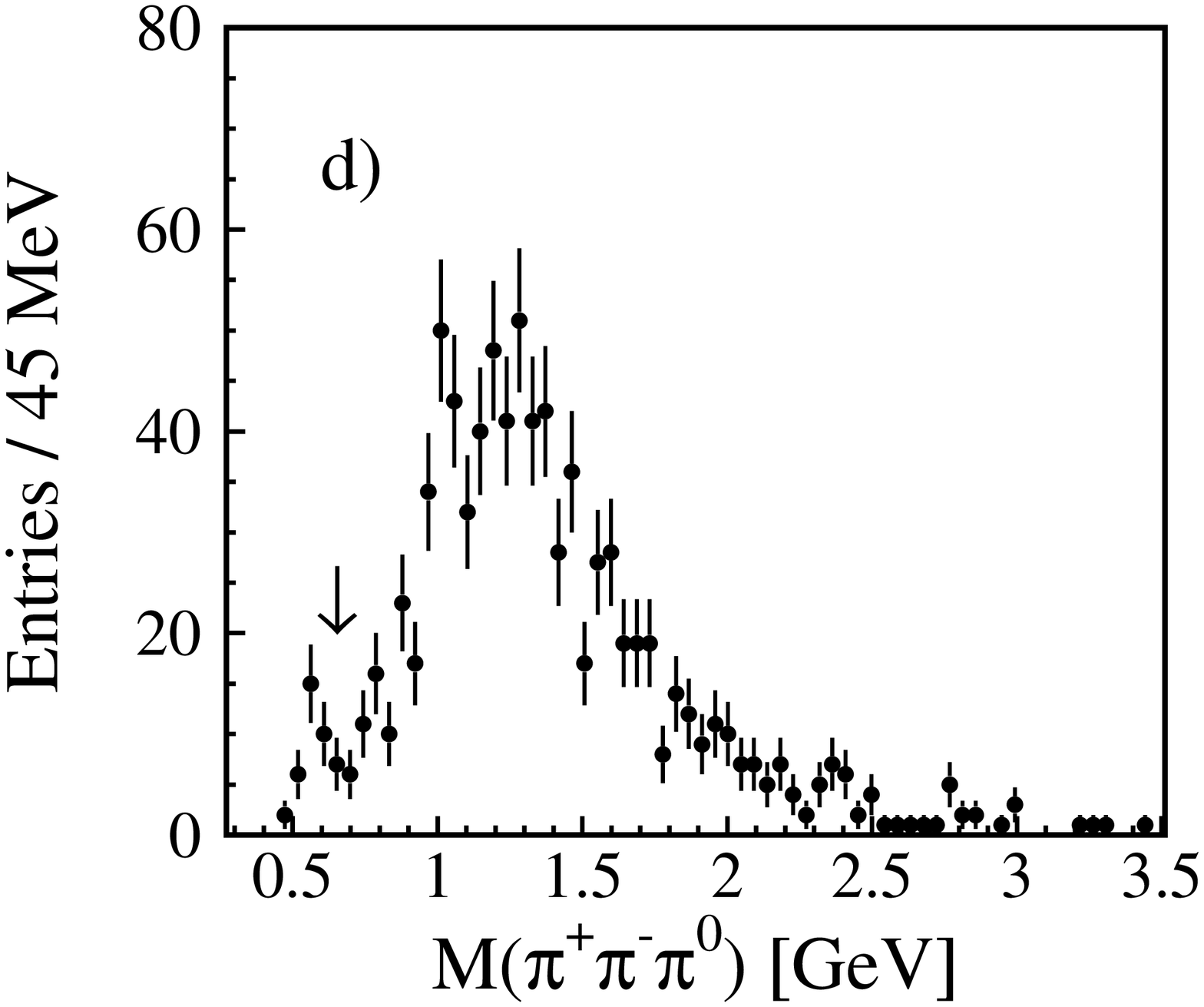,width=0.49\textwidth}}
  \end{center}
  \caption[]{Distributions for $\pipi\pizpiz$ candidates.
             a) Two-photon invariant mass  for the selected $\piz$'s  (two entries per event);
             b) azimuthal angle, $\phi_{tag}$, of the
             tagged electron 
             for tags in the inner side of the LEP ring (in) and,
             folded over it, for tags in the outer side of
             the LEP ring (out);
             c) acoplanarity angle, $\phi_{aco}$,
             between the tagged electron and the $\pipi\pizpiz$ system and            
             d) mass of the $\pipi\piz$ system (two entries per event).
	     The data  are compared to the 
	     four-pion Monte Carlo. The estimated background is indicated by
	     the hatched histograms.  
             The arrows indicate the selection cuts.
            }
\label{fig:fig1} 
\end{figure}

Single-tagged events are selected by requiring just one electromagnetic cluster with 
energy greater then  $50\% $ of the beam energy  reconstructed in the VSAT.
The event candidates must have exactly two tracks with zero total
charge.  The tracks must come from the interaction vertex, have
transverse momentum greater than $100 \MeV{}$ and an energy loss in
the tracking chamber compatible with the pion hypothesis.
The selected events should contain a $\pizpiz$ pair, therefore we consider
event candidates that have four or five photons, identified as isolated clusters
in the electromagnetic calorimeter. Photons  having energies greater than $60 \MeV$ 
are  paired to  reconstruct neutral pions, which are required to be in the 
mass window $100 \MeV \le  { M(\gamma\gamma)} \le  170 \MeV$, 
as shown in  Figure~\ref{fig:fig1}a. The mass of a $\piz$ candidate is 
constrained to the nominal value by a \mbox{1-C} kinematic fit.
If more than one $\pizpiz$ combination exists,
the one with the smallest  $\chi^2$ sum of the fits 
is taken.
To make the selection robust against instrumental noise and backgrounds 
and to reduce the sensitivity to the Monte Carlo  simulation of fake photons,
we retain events with one additional photon, not used in the $\pizpiz$ pair, 
if the photon energy is less than 300 \MeV{} and 
does not exceed $10\%$ of the energy of the $\pizpiz$ pair.

The transverse momentum squared, $\ptt$, of the four-pion system is used to measure the $\q$ of
the event and is required to be in the range $0.2 \GeV^2 -  0.85\GeV^2$.  
For selection of an exclusive final state,
the acoplanarity angle, $\phi_{aco}$, calculated from the difference
between the azimuthal angle of the tagged electron, $\phi_{tag}$, shown in Figure~\ref{fig:fig1}b,
and the azimuthal angle of the four-pion system, is required to be less than 0.4 rad, 
as shown in Figure~\ref{fig:fig1}c.
The data contain a contribution from  $\eta$ production, as visible in the
the $\pipi\piz$ mass spectrum, shown in Figure~\ref{fig:fig1}d. 
This background is removed by requiring  
$M( \pipi\piz) > 0.65 \GeV$.

\begin{figure}[p]
\begin{center}
{\epsfig{file=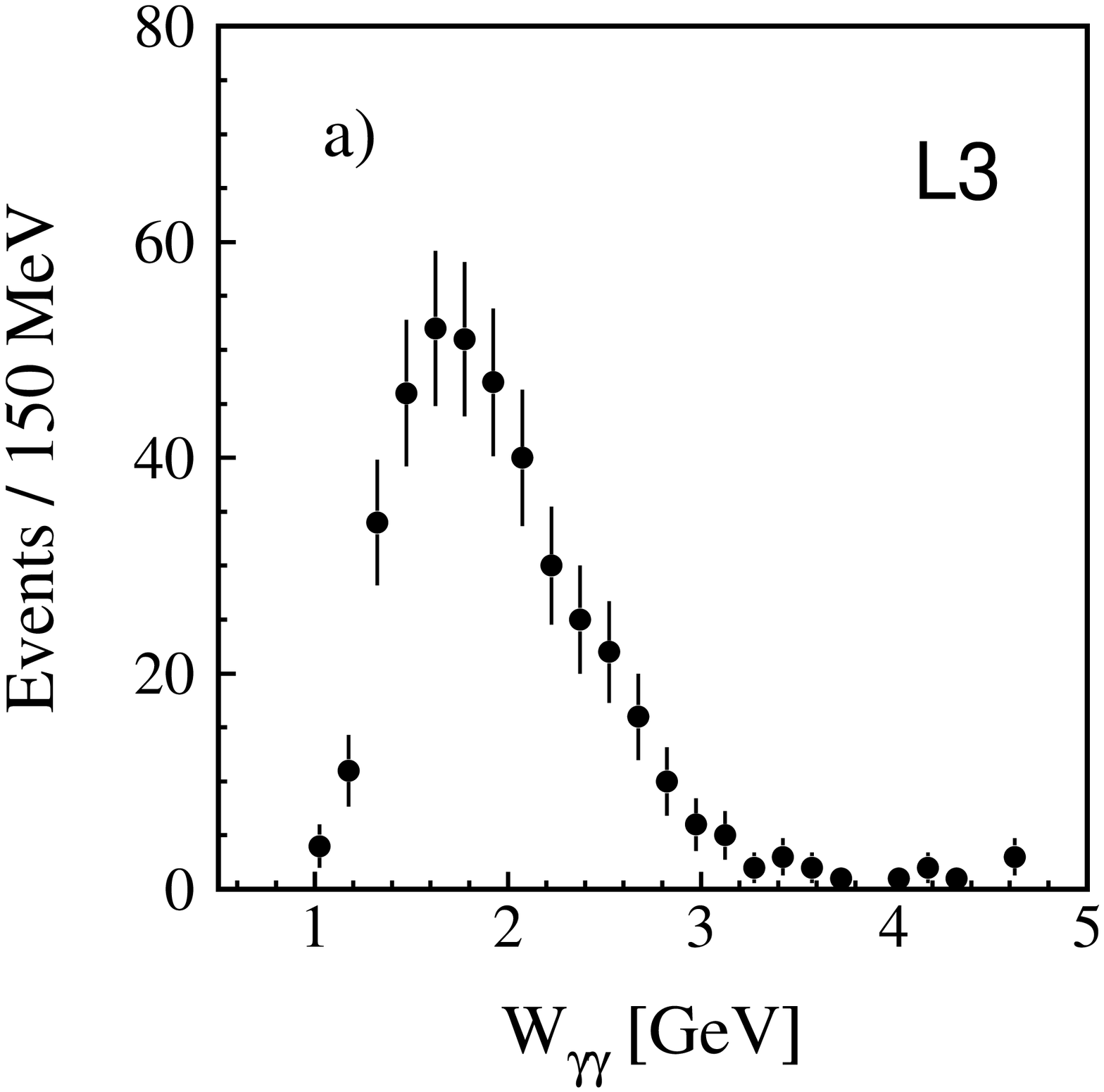,width=0.49\linewidth}}
{\epsfig{file=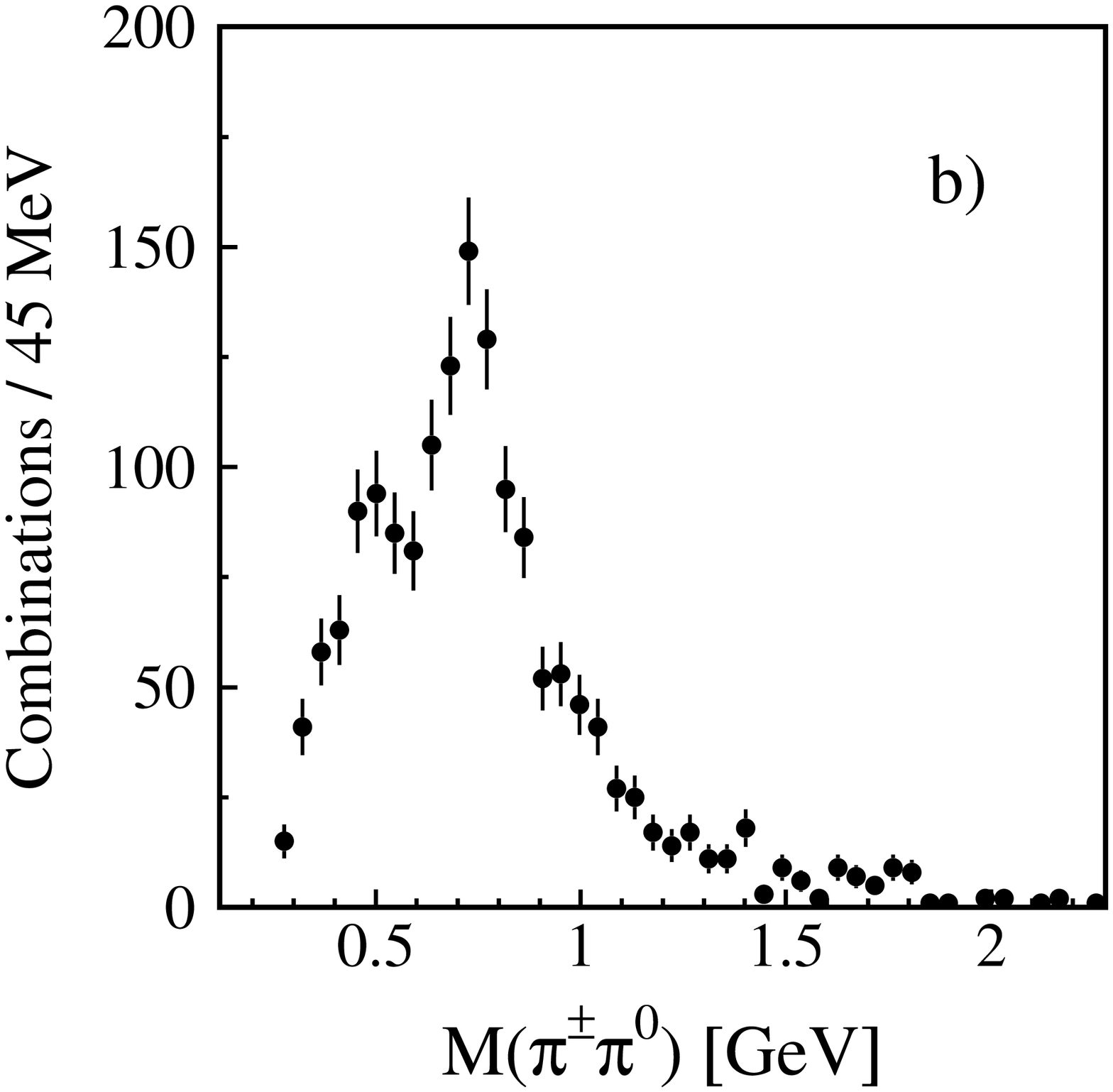,width=0.49\linewidth}}
{\epsfig{file=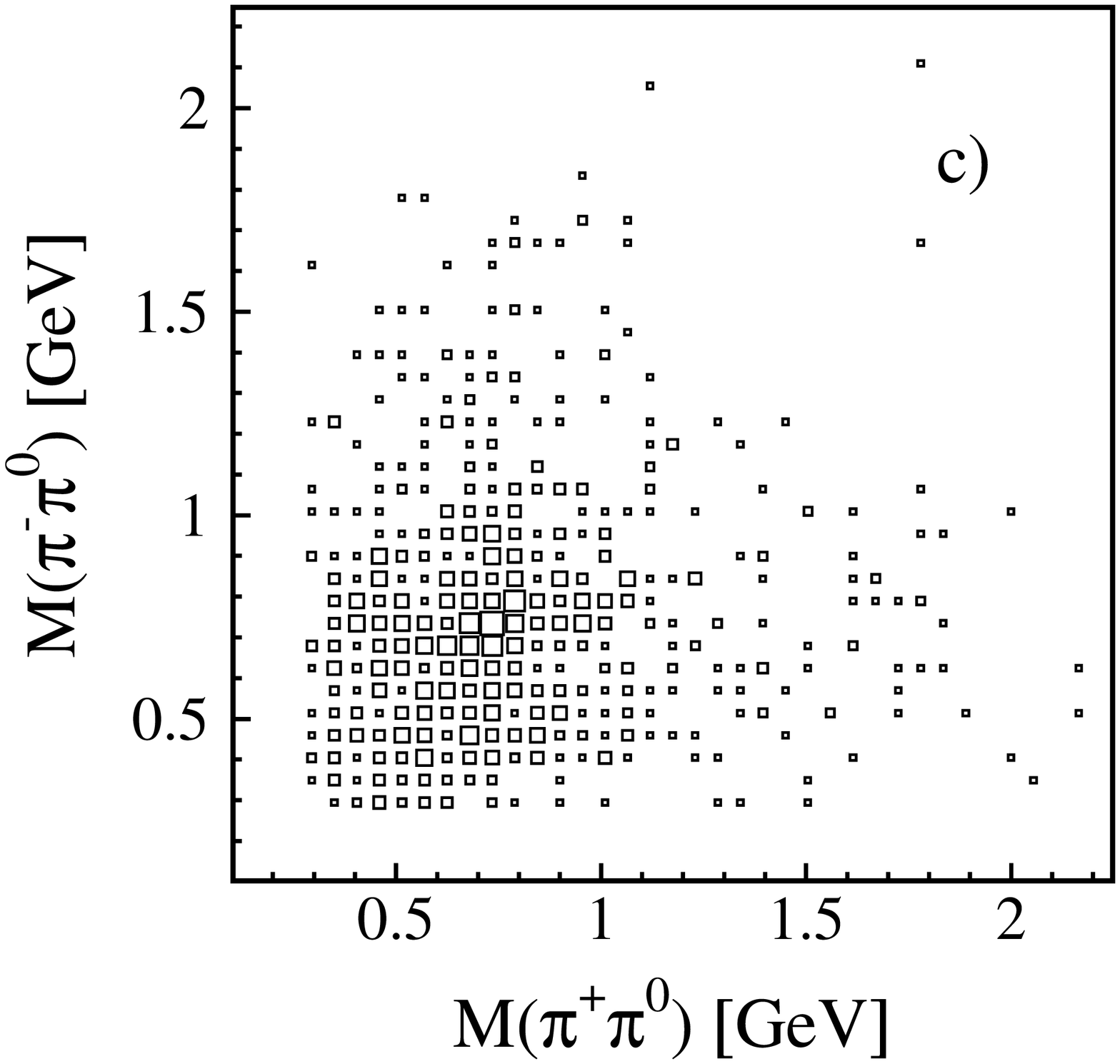,width=0.49\linewidth}}
 {\epsfig{file=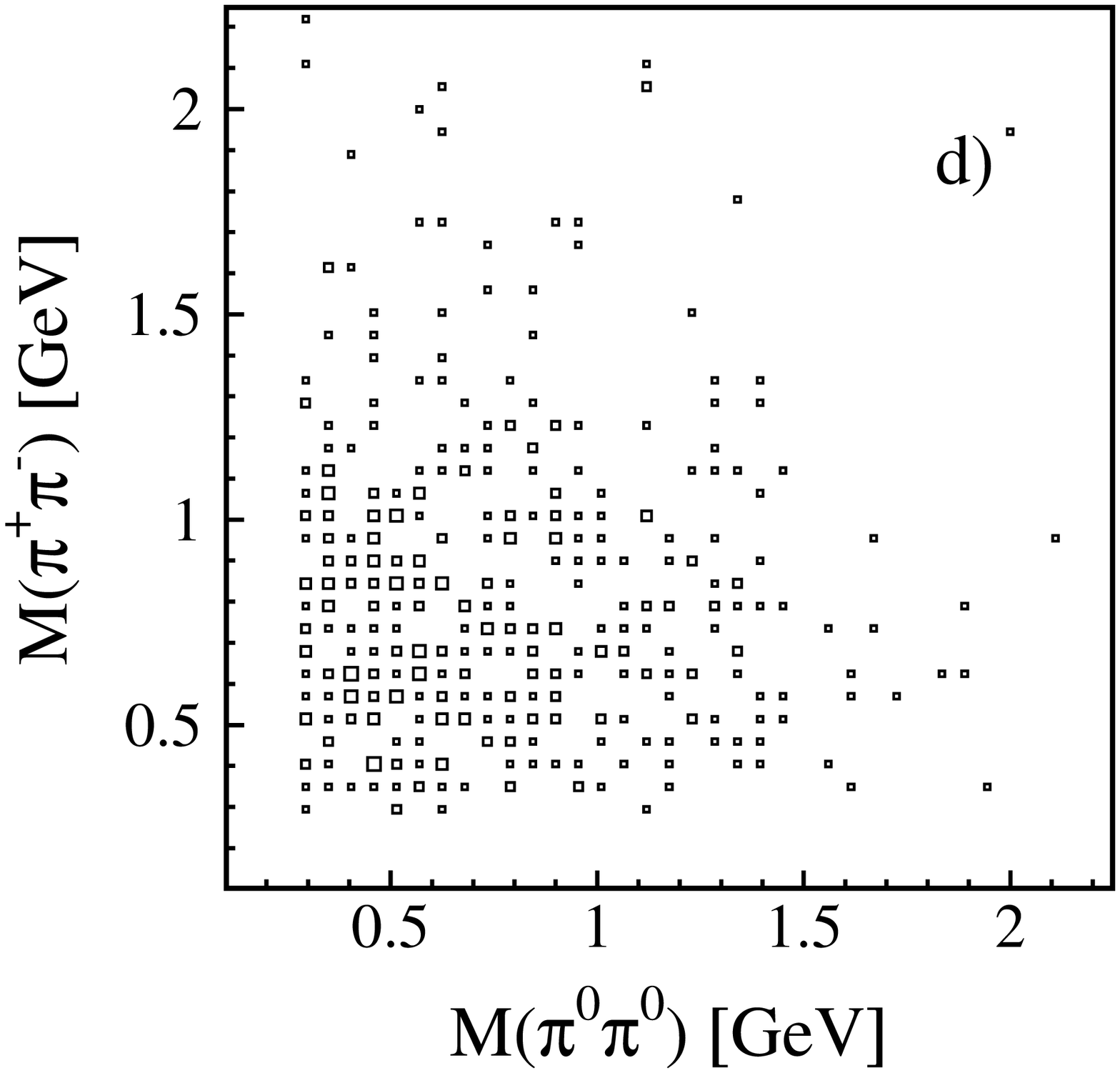,width=0.49\linewidth}}
\end{center}

  \caption[]{Mass distributions for the selected events:
             a) the four-pion system, $\mgg$;
             b) the $\pi^\pm \piz$ combinations (four entries per event);
             c) correlation between  the $\pi^-\piz$ and  $\pi^+\piz$
                 pairs (two entries per event) and
             d) correlation between  the $\pipi$ and  $\pizpiz$ pairs.
             The two-dimensional distributions have a bin width of
             $55 \times 55 \MeV^2$, the size of the boxes is proportional to the number
             of entries and both plots have the same vertical scale.
 
           }
\label{fig:fig2}
\end{figure}

After all cuts, 414 events are retained. Their four-pion mass
spectrum is shown in Figure~\ref{fig:fig2}a. The region $1.1 \GeV  \le   \mgg \le  3 \GeV$ 
 is populated by 387 events, which are used for
the cross section determination. 
A strong signal from $\rho ^\pm$ production   is observed in
 the $ M (\pi^\pm\piz )$ spectrum,
 shown in Figure~\ref{fig:fig2}b.
The clustering of entries at the  crossing of the $\rho^\pm$
mass bands in the correlation plot of the masses of the $\pi^\pm\piz$ combinations,
shown in Figure~\ref{fig:fig2}c, gives evidence for a signal from $\rocroc$ intermediate states.
No structure is observed in the correlation plot of the masses of the $\pipi$ 
and $\pizpiz$ combinations, shown in Figure~\ref{fig:fig2}d.
We also inspected the two- and three-pion mass distributions, 
shown in  Figure~\ref{fig:fig3},
for production of higher-mass resonances.
The only statistically-significant signal is from the $a_2^\pm(1320)$ 
state in the $\pi^\pm\pizpiz$  mass spectrum, 
as seen in Figure~\ref{fig:fig3}f.

\section {Data Analysis}
\subsection {Monte Carlo Modelling}

To estimate the number of $\rocroc$ events in the selected four-pion data sample,
we consider non-interfering contributions from the processes:
\begin{eqnarray}
\label{eq:eqn04}
&&  \gamgam \to \rocroc;   \nonumber\\ 
&&  \gamgam \to  \roc\pi^\mp \pi^0;  \\
&&  \gamgam \to a_2^\pm(1320) \pi^\mp \,; \nonumber\\
&&  \gamgam \to \pipi \pizpiz ,\, \mathrm{non \!-\! resonant}.  \nonumber 
\end{eqnarray}
About 40 million Monte Carlo events of the processes (\ref{eq:eqn04}) 
are generated with the EGPC~\cite{LINDE} program, which uses the
luminosity function from Reference~\citen{BUDNEV}.
Particle production and decay is uniform in phase-space.
The generated events are passed through the full
L3 detector simulation using the GEANT \cite{geant} and GHEISHA
\cite{GHEISHA} programs and processed in the same way as the data,
reproducing the detector behaviour as monitored in the different
data-taking periods.

For acceptance calculations,  Monte Carlo events are assigned a
$\q$-dependent weight, evaluated using the GVDM
form-factor~\cite{GVDM}  for both interacting photons. 
The detection efficiencies of the process (\ref{eq:eqn01}) are
listed in Tables~\ref{tbl:xsectq2} and~\ref{tbl:xsectwgg} for bins in $\q$ and $\mgg$.
The efficiencies for the four-pion final states of all the processes
(\ref{eq:eqn04}) are of similar magnitude.

\subsection {Background Estimation}

The contribution to the selected events from $\EE$ annihilation  
and from the  process
$\EE\rightarrow\EE\tau^+\tau^-$ is negligible.
Random coincidences with off-momentum beam electrons, 
which give  signals in the VSAT, are a source of background. The flux of these particles is  
dominantly on the outer side of the LEP ring. Therefore, this background 
would cause an excess in the number of events having a tag on the outer side 
of the accelerator ring, $N_{out}$, with respect to the inner side, $N_{in}$. 
In the selected data, the ratio $N_{out}/N_{in} = 1.04 \pm 0.10$ is close to unity, 
indicating that this background is small. 
This conclusion is corroborated by the good agreement observed between the 
$\phi_{tag}$  distribution of the selected data and 
Monte Carlo event samples, shown in Figure~\ref{fig:fig1}b. 

Two sources of background remain. The first is  partially-reconstructed events from two-photon
interactions with higher particle multiplicities, when tracks or photons escape detection. 
The second is signal events with one or more photons   
substituted by photon candidates due to noise.
To estimate  the accepted  background we use background-like 
event samples extracted from the experimental data. 
The first background is modelled with selected  $\pi^\pm \pi^\pm \pizpiz$ events, 
in which at least two charged particles have not been detected and by
$\pipi\pizpiz\piz$ events in which one $\piz$ is excluded from consideration.
An event-mixing technique is employed in order to reproduce events from the second
background: one or two photons forming a $\piz$ are excluded  from a selected 
event and replaced by photons from another data event.
The  $\phi_{aco}$ distributions of the  back\-ground-like data samples, passing the selection,  
are combined with the distribution of selected $\pipi\pizpiz$ Monte Carlo
events so as to reproduce the $\phi_{aco}$ distribution observed in the data,  
 as shown in Figures~\ref{fig:fig1}c. The estimated background levels 
are listed in Tables~\ref{tbl:xsectq2} and~\ref{tbl:xsectwgg}.
As data samples are used in the background estimation,
 they contain also a fraction of events with fake tags and thus take into account the
effect of this background.

\subsection {Fit Method}

In order to determine the differential $\rocroc$ production rate, a
maximum likelihood fit of the data to a sum of Monte Carlo samples of the processes
(\ref{eq:eqn04}) is performed in intervals of $\q$ and $\mgg$ using 
a box method\cite{L3paper269, L3paper287, L3paper292, BOXMETHOD}. 
The inputs to the fit are the six two-pion masses in an event, namely 
the four combinations  $\pi^\pm\piz$ and the two combinations
$\pipi$ and $\pizpiz$. They provide a complete description of a four-pion event 
in our model of isotropic production and phase space decay.

The analysis procedure is optimised for deriving the $\rocroc$ contribution 
and only the $\rocroc$ content and the sum of the rest 
of the contributing processes, denoted as ``other $4\pi$'', are considered for cross section measurements.
To check the quality of the fit, the two- and three-pion mass distributions 
of the data are compared in Figure~\ref{fig:fig3} with those of a mixture of Monte Carlo event samples
from the processes (\ref{eq:eqn04}), in proportions determined by the fit.  
The observed experimental distributions are reasonably well described by the Monte Carlo model.

\section {Results}

The cross sections of the process $\EE \to\EE\rocroc$ 
  in bins of $\q$ and $\mgg$, $\mathrm{\Delta \sigma_{ee}}$, 
  are listed in Tables~\ref{tbl:xsectq2} and~\ref{tbl:xsectwgg}.
The statistical uncertainties, also listed in   Tables~\ref{tbl:xsectq2}
and~\ref{tbl:xsectwgg}, are those of the fit.  
The differential cross section, $d \sigma_{\mathrm{ee}} / d \q$,
derived from $\mathrm{\Delta \sigma_{ee}}$, is listed in
Table~\ref{tbl:xsectq2}.  When evaluating the differential cross
section, a correction based on the $\q$-dependence of the $\rocroc$
Monte Carlo sample is applied, so as to assign the cross section value
to the centre of the corresponding $\q$-bin~\cite{BIN}.
 
To evaluate the cross section, $\sigma_{\gamma\gamma}$, of the process
$\gamgam \to \rocroc$, the integral of the transverse photon luminosity
function, $L_{TT}$, is computed for each $\q$ and $\mgg$ bin using the
program GALUGA \cite{GALUGA}, which performs ${\cal O}(\alpha^4)$ QED
calculations. The same procedure was
used in our  previous studies~\cite{L3paper269, L3paper292,L3paper287}.
The cross section $\sigma_{\gamma\gamma}$ is derived
from the measured cross section  using
the relation
$\sigma_{\gamma\gamma} = \mathrm{\Delta \sigma_{ee}} /  L_{TT}$.  
Thus, $\sigma_{\gamma\gamma}$ represents an
effective cross section containing contributions from both transverse
and longitudinal photon polarisations.  The cross section of the
process $ \gamgam \to \rocroc$ is listed in Table \ref{tbl:xsectq2} as a
function of $\q$ and in Table~\ref{tbl:xsectwgg} as a function of
$\mgg$.  The sum of the cross sections of the other contributing
processes is also given in Tables~\ref{tbl:xsectq2}
and~\ref{tbl:xsectwgg}.

Several sources of systematic uncertainty are considered. The
contribution of the selection procedure is in the range $12\% - 18\%$;  
Monte Carlo statistics in the range \mbox{$1.3\% - 2.1\%$}; 
the fit procedure in the range $11\% - 20\%$.
Half of the changes of the acceptance when no form factor re-weighting 
of the Monte Carlo events is performed is considered as model uncertainty.
It is in the range \mbox{$0.5\% - 5\%$}.
The background correction procedure introduces systematic
uncertainties in the range \mbox{$2\% - 6\%$}.
All contributions are added in quadrature to obtain the systematic
uncertainties, quoted in Tables~\ref{tbl:xsectq2} and~\ref{tbl:xsectwgg}. 
Finally, a normalization uncertainty of 5\% accounts for the uncertainty of the 
trigger efficiency determination.

\section {Study of {\boldmath$\gamgam \to  \rho\rho $} Process}
Combining the present results  with the L3 data 
on  $\rho\rho$ production from References \citen{L3paper269, L3paper287,  L3paper292},
we compare the $\rocroc$ to the $\roro$ channels and their evolution as a function of 
$\q$.
The cross section of the process $ \gamgam \to \rho \rho$ is plotted in Figure~\ref{fig:fig5} 
as a function of $\mgg$.
For $\mgg  \le 2.1 \GeV $ and  $0.2 \GeV^2  \le  \q  \le 0.85\GeV^2$
there is a clear enhancement of $\roro$ production
relative to $\rocroc$. This is similar to what was observed 
at $\q \approx 0$~\cite{ARGUS, TASSO}, but 
 in contrast with the high $\q$-region, where both cross sections have similar
dependence on  $\mgg$ and 
the $\rocroc$ is systematically higher than the $\roro$.
These differences  are clearly seen in the ratio  
$ \mathrm{R} = \sum{\Delta \sigma_{ee}(\rocroc)} / \sum{\Delta \sigma_{ee}(\roro)}$
where the sum is for  the region $ 1.1 \GeV \le  \mgg  \le 2.1 \GeV $.
In the domain $ 0.20 \GeV ^2 \le \q \le 0.85 \GeV^2$, we measure
$  \mathrm{R} = 0.62 \pm 0.10\,(\rm stat.)  \pm 0.09\,(\rm syst.)$,   
a value that can only be explained  by the presence of
an isospin \mbox{I = 2} intermediate state or by a mixture of different 
states~\cite{ATTEMPTS,rosner}.  
The value of this ratio for  $ 1.2 \GeV ^2 \le \q \le 8.5 \GeV^2$ is 
$  \mathrm{R}  = 1.81 \pm 0.47\,(\rm stat.)  \pm 0.22\,(\rm syst.)$ 
\cite{L3paper287},   close to the factor 2, expected for an isospin \mbox{I = 0} state. 
Such variation suggests  different $\rho$-pair production mechanisms 
at low and high  $\q$.

The differential cross section $d \sigma_{\mathrm{ee}} / d \q$ of the reaction 
$\EE \to\EE\rho\rho$ is shown in
Figure~\ref{fig:fig6}a.
The  L3 measurements span   a $\q$-region of two orders of magnitude,
over which the differential  cross sections show a monotonic fall of
more than four orders of magnitude.
The $ \rho\rho$  data are fitted to a form~\cite{DIEHL} expected from QCD-based
calculations\cite{DIEHLPAP}:
\begin{equation}
\label{eq:eqn11}
{ \mathrm{d} \sigma_{\mathrm{ee}} \over \mathrm{d} \q }\sim \frac{1} { Q^n
(\q + < \mgg >^2)^2} \, ,
\end{equation}
where $n$ is a constant and $< \mgg >$ is the average $\mgg$ value, 
1.9 \GeV\ for this measurement. Although this formula is expected to be valid
only for $\q \gg \mgg$, we find it provides a good parametrisation of
the $\q$-evolution of the $\rho\rho$  data.
A fit to the $\rocroc$ data  finds
 an exponent $ n = 2.3 \pm 0.2$ with $\chi^2 /d.o.f. = 1.4/7$.
 A value $ n = 2.9 \pm 0.1$ was found for $\roro$ 
 with $\chi^2 /d.o.f. = 6.9/10$~\cite{L3paper292}.
Only the statistical uncertainties are considered in the fits.
The results of the fits are shown  in Figure~\ref{fig:fig6}a. 
The fits indicate a cross-over of the  
differential cross sections of 
$\rocroc$ and $\roro$ production in the vicinity of $\q \approx 1 \GeV^2$. 

The measured cross section of the process $ \gamgam \to \rho \rho$ as a
function of $\q$ is shown in Figure~\ref{fig:fig6}b. 
The change of the relative magnitude of  $\rocroc$ and $\roro$ production 
is clearly visible when comparing the low-  and the high-$\q$ regions. 
A parametrisation, based on the GVDM model~\cite{GVDM}: 
\begin{equation} 
\label{eq:eqn12}
\sigma _{\gamma\gamma} (W_{\gamma\gamma},\q)= \sigma _{\gamma\gamma} (W_{\gamma\gamma}) F(\q),\;\;\;
\mathrm{and} \;\;\;
 F(\q) = \sum_{V = \rho , \omega , \phi } r_V \frac{1 + \q /4 m_V^2}{(1 + \q /m_V^2)^2} + 
\frac{0.22} {1 + Q^2 /m_0^2} , \\
\end{equation}
 with  $ r_{\rho} = 0.65 , r_{\omega} = 0.08 ,r_{\phi} = 0.05$  and  $m_0 = 1.4 \GeV $   
 reproduces well the $\q$-dependence 
of the $\roro$ data, as shown in Reference~\citen{L3paper292} and indicated
by the line in Figure~\ref{fig:fig6}b.
The fit finds a cross section of $ 13.6 \pm 0.7 $ nb 
for the $\mgg$ region $1.1 \GeV  \le   \mgg \le  3 \GeV $ at  $\q = 0$.
The  $\q$-evolution of $\rocroc$ data  cannot be satisfactorily described
by this form. In addition, as shown in Figure~\ref{fig:fig6}b,
the $\roro$ data cannot be described by the much steeper $\q$-fall of 
a $\rho$-pole parametrisation~\cite{L3paper292};
the same is true for the  $\rocroc$ cross section  since it is decreasing with $\q$ less steeply
than the $\roro$ one.

\section {Conclusions}

We have performed the first measurement of exclusive $\rocroc$ production in 
mid-virtuality two-photon collisions. These  results complement 
 previous L3 measurements of exclusive 
 $\rocroc$ and $\roro$ production and allow to follow the evolution of $\rho\rho$ cross sections 
over a $\q$-region of two orders of magnitude.

A QCD-based form, derived for the description of the differential cross-section
$d \sigma_{\mathrm{ee}} / d \q$ of the process $\EE \to\EE\rho\rho$ 
at high $\q$, is found to provide a good parametrisation of
the $\q$-evolution of the $\rho\rho$  data in the entire interval 
$0.2 \GeV^2 \le \q \le 30 \GeV^2$,
over which the differential  cross sections show a monotonic decrease 
of more than four orders of magnitude, for $1.1 \GeV  \le   \mgg \le  3 \GeV $.

The  $\q$-dependence of the cross section of the process $ \gamgam \to \roro$
is well reproduced by  a parametrisation  based on the GVDM model over the entire $\q$-region. 
On the other hand,
   the  $\rocroc$ data  cannot be satisfactorily described
by such a parametrisation. 
 A $\rho$-pole description of the $\q$-dependence for both 
$\roro$ and $\rocroc$ data is excluded. 

The relative magnitude of  $\rocroc$ and $\roro$ production changes in the vicinity 
of $\q \approx 1 \GeV^2$, suggesting  different $\rho$-pair production mechanisms 
at low and high  $\q$.


%
%
%
%
\newpage
\typeout{   }     
\typeout{Using author list for paper 287 -  }
\typeout{$Modified: Jul 15 2001 by smele $}
\typeout{!!!!  This should only be used with document option a4p!!!!}
\typeout{   }
%
%
%
%
%
%

\newcount\tutecount  \tutecount=0
\def\tutenum#1{\global\advance\tutecount by 1 \xdef#1{\the\tutecount}}
\def\tute#1{$^{#1}$}
\tutenum\aachen            
\tutenum\nikhef            
\tutenum\mich              
\tutenum\lapp              
\tutenum\basel             
\tutenum\lsu               
\tutenum\beijing           
\tutenum\bologna           
\tutenum\tata              
\tutenum\ne                
\tutenum\bucharest         
\tutenum\budapest          
\tutenum\mit               
\tutenum\panjab            
\tutenum\debrecen          
\tutenum\dublin            
\tutenum\florence          
\tutenum\cern              
\tutenum\wl                
\tutenum\geneva            
\tutenum\hamburg           
\tutenum\hefei             
\tutenum\lausanne          
\tutenum\lyon              
\tutenum\madrid            
\tutenum\florida           
\tutenum\milan             
\tutenum\moscow            
\tutenum\naples            
\tutenum\cyprus            
\tutenum\nymegen           
\tutenum\caltech           
\tutenum\perugia           
\tutenum\peters            
\tutenum\cmu               
\tutenum\potenza           
\tutenum\prince            
\tutenum\riverside         
\tutenum\rome              
\tutenum\salerno           
\tutenum\ucsd              
\tutenum\sofia             
\tutenum\korea             
\tutenum\taiwan            
\tutenum\tsinghua          
\tutenum\purdue            
\tutenum\psinst            
\tutenum\zeuthen           
\tutenum\eth               

{
\parskip=0pt
\noindent
{\bf The L3 Collaboration:}
\ifx\selectfont\undefined
 \baselineskip=10.8pt
 \baselineskip\baselinestretch\baselineskip
 \normalbaselineskip\baselineskip
 \ixpt
\else
 \fontsize{9}{10.8pt}\selectfont
\fi
\medskip
\tolerance=10000
\hbadness=5000
\raggedright
\hsize=162truemm\hoffset=0mm
\def\r{\rlap,}
\noindent

P.Achard\r\tute\geneva\ 
O.Adriani\r\tute{\florence}\ 
M.Aguilar-Benitez\r\tute\madrid\ 
J.Alcaraz\r\tute{\madrid}\ 
G.Alemanni\r\tute\lausanne\
J.Allaby\r\tute\cern\
A.Aloisio\r\tute\naples\ 
M.G.Alviggi\r\tute\naples\
H.Anderhub\r\tute\eth\ 
V.P.Andreev\r\tute{\lsu,\peters}\
% ...
\newpage
\begin{list}{A}{\itemsep=0pt plus 0pt minus 0pt\parsep=0pt plus 0pt minus 0pt
                \topsep=0pt plus 0pt minus 0pt}
\item[\aachen]
 III. Physikalisches Institut, RWTH, D-52056 Aachen, Germany$^{\S}$
\item[\nikhef] National Institute for High Energy Physics, NIKHEF, 
     and University of Amsterdam, NL-1009 DB Amsterdam, The Netherlands
\item[\mich] University of Michigan, Ann Arbor, MI 48109, USA
\item[\lapp] Laboratoire d'Annecy-le-Vieux de Physique des Particules, 
     LAPP,IN2P3-CNRS, BP 110, F-74941 Annecy-le-Vieux CEDEX, France
\item[\basel] Institute of Physics, University of Basel, CH-4056 Basel,
     Switzerland
\item[\lsu] Louisiana State University, Baton Rouge, LA 70803, USA
\item[\beijing] Institute of High Energy Physics, IHEP, 
  100039 Beijing, China$^{\triangle}$ 
\item[\bologna] University of Bologna and INFN-Sezione di Bologna, 
     I-40126 Bologna, Italy
\item[\tata] Tata Institute of Fundamental Research, Mumbai (Bombay) 400 005, India
\item[\ne] Northeastern University, Boston, MA 02115, USA
\item[\bucharest] Institute of Atomic Physics and University of Bucharest,
     R-76900 Bucharest, Romania
\item[\budapest] Central Research Institute for Physics of the 
     Hungarian Academy of Sciences, H-1525 Budapest 114, Hungary$^{\ddag}$
\item[\mit] Massachusetts Institute of Technology, Cambridge, MA 02139, USA
\item[\panjab] Panjab University, Chandigarh 160 014, India
\item[\debrecen] KLTE-ATOMKI, H-4010 Debrecen, Hungary$^\P$
\item[\dublin] Department of Experimental Physics,
  University College Dublin, Belfield, Dublin 4, Ireland
\item[\florence] INFN Sezione di Firenze and University of Florence, 
     I-50125 Florence, Italy
\item[\cern] European Laboratory for Particle Physics, CERN, 
     CH-1211 Geneva 23, Switzerland
\item[\wl] World Laboratory, FBLJA  Project, CH-1211 Geneva 23, Switzerland
\item[\geneva] University of Geneva, CH-1211 Geneva 4, Switzerland
\item[\hamburg] University of Hamburg, D-22761 Hamburg, Germany
\item[\hefei] Chinese University of Science and Technology, USTC,
      Hefei, Anhui 230 029, China$^{\triangle}$
\item[\lausanne] University of Lausanne, CH-1015 Lausanne, Switzerland
\item[\lyon] Institut de Physique Nucl\'eaire de Lyon, 
     IN2P3-CNRS,Universit\'e Claude Bernard, 
     F-69622 Villeurbanne, France
\item[\madrid] Centro de Investigaciones Energ{\'e}ticas, 
     Medioambientales y Tecnol\'ogicas, CIEMAT, E-28040 Madrid,
     Spain${\flat}$ 
\item[\florida] Florida Institute of Technology, Melbourne, FL 32901, USA
\item[\milan] INFN-Sezione di Milano, I-20133 Milan, Italy
\item[\moscow] Institute of Theoretical and Experimental Physics, ITEP, 
     Moscow, Russia
\item[\naples] INFN-Sezione di Napoli and University of Naples, 
     I-80125 Naples, Italy
\item[\cyprus] Department of Physics, University of Cyprus,
     Nicosia, Cyprus
\item[\nymegen] Radboud University and NIKHEF, 
     NL-6525 ED Nijmegen, The Netherlands
\item[\caltech] California Institute of Technology, Pasadena, CA 91125, USA
\item[\perugia] INFN-Sezione di Perugia and Universit\`a Degli 
     Studi di Perugia, I-06100 Perugia, Italy   
\item[\peters] Nuclear Physics Institute, St. Petersburg, Russia
\item[\cmu] Carnegie Mellon University, Pittsburgh, PA 15213, USA
\item[\potenza] INFN-Sezione di Napoli and University of Potenza, 
     I-85100 Potenza, Italy
\item[\prince] Princeton University, Princeton, NJ 08544, USA
\item[\riverside] University of Californa, Riverside, CA 92521, USA
\item[\rome] INFN-Sezione di Roma and University of Rome, ``La Sapienza",
     I-00185 Rome, Italy
\item[\salerno] University and INFN, Salerno, I-84100 Salerno, Italy
\item[\ucsd] University of California, San Diego, CA 92093, USA
\item[\sofia] Bulgarian Academy of Sciences, Central Lab.~of 
     Mechatronics and Instrumentation, BU-1113 Sofia, Bulgaria
\item[\korea]  The Center for High Energy Physics, 
     Kyungpook National University, 702-701 Taegu, Republic of Korea
\item[\taiwan] National Central University, Chung-Li, Taiwan, China
\item[\tsinghua] Department of Physics, National Tsing Hua University,
      Taiwan, China
\item[\purdue] Purdue University, West Lafayette, IN 47907, USA
\item[\psinst] Paul Scherrer Institut, PSI, CH-5232 Villigen, Switzerland
\item[\zeuthen] DESY, D-15738 Zeuthen, Germany
\item[\eth] Eidgen\"ossische Technische Hochschule, ETH Z\"urich,
     CH-8093 Z\"urich, Switzerland
\item[\S]  Supported by the German Bundesministerium 
        f\"ur Bildung, Wissenschaft, Forschung und Technologie.
\item[\ddag] Supported by the Hungarian OTKA fund under contract
numbers T019181, F023259 and T037350.
\item[\P] Also supported by the Hungarian OTKA fund under contract
  number T026178.
\item[$\flat$] Supported also by the Comisi\'on Interministerial de Ciencia y 
        Tecnolog{\'\i}a.
\item[$\sharp$] Also supported by CONICET and Universidad Nacional de La Plata,
        CC 67, 1900 La Plata, Argentina.
\item[$\triangle$] Supported by the National Natural Science
  Foundation of China.
\end{list}
}
\vfill


\newpage

\begin{table}[htb]
\begin{center}
 \begin{sideways}
 \begin{minipage}[b]{\textheight}
\begin{center}
\begin{tabular}{|c|c|c|c|c|c|c|}
\hline
$ \q$ range &  $\varepsilon$ & $\mathit{Bg}$ & $\Delta \sigma_{ee}$ [pb] & $ d\,\sigma_{ee}/d\,\q$ [pb/$\rm Ge\kern -0.1em V^2]$ & 
$ \sigma_{\gamma\gamma}$ [nb]  & $ \sigma_{\gamma\gamma}$ [nb] \\
 $[\rm Ge\kern -0.1em V^2 ]$ & [ \% ] & [ \% ] & $\rocroc$ & $\rocroc$ &  $\rocroc$ & other $4\pi$  \\ \hline

0.20 -- 0.28 & 0.8 & 14 & $   7.4 \pm 2.4  \pm 1.9 $ & $ 92 \pm 29\phantom{.0} \pm 23\phantom{.0} $  & $ 5.7 \pm 1.8 \pm 1.4 $ & $ 10.9 \pm 2.2 \pm 1.5 $ \\ \hline
0.28 -- 0.40 & 1.2 & 14 & $   5.7 \pm 1.8  \pm 1.3 $ & $ 47 \pm 15\phantom{.0} \pm 10\phantom{.0} $  & $ 4.3 \pm 1.4 \pm 1.0 $ & $ 12.2 \pm 1.8 \pm 1.4 $ \\ \hline
0.40 -- 0.55 & 1.1 & 15 & $   5.6 \pm 1.6  \pm 1.1 $ & $ 37 \pm 11\phantom{.0} \pm \phantom{0}7.3$  & $ 4.9 \pm 1.4 \pm 1.0 $ & $ 13.3 \pm 2.0 \pm 1.8 $ \\ \hline
0.55 -- 0.85 & 0.7 & 18 & $   7.7 \pm 2.5  \pm 2.0 $ & $ 25 \pm \phantom{0}8.2 \pm \phantom{0}6.5$& $ 5.3 \pm 1.7 \pm 1.4 $ & $ 12.1 \pm 2.2 \pm 1.9 $ \\ \hline
\end{tabular}

\caption{Detection efficiencies, $\varepsilon$,  background fractions, $\mathit{Bg}$,
         and  cross sections 
         of the reactions $ \EE \to \EE \rocroc$, $\gamgam \to \rocroc$ 
	        and  of the sum of the
         rest of the contributing processes, "other $4\pi$",
         as a function of  $\q$ for $1.1\GeV \le \mgg \le 3\GeV$. 
         The values of the differential cross sections are   corrected to the centre of each bin.
         The first uncertainties are statistical, the second systematic.
	 An overall normalization uncertainty of $5 \% $ for the trigger is not included.
         }
\label{tbl:xsectq2}
\end{center}
\end{minipage}
\end{sideways}
\end{center}
\end{table}

\begin{table*}[ht]
\begin{center}
\begin{tabular}{|c|c|c|c|c|c|}
\hline
$  \mgg$-range & $\varepsilon$&  $\mathit{Bg}$ &  $ \Delta \sigma_{ee}$ [pb] & 
$ \sigma_{\gamma\gamma}$ [nb]  & $ \sigma_{\gamma\gamma}$ [nb] \\
$[\rm Ge\kern -0.1em V ]$ & [ \% ]  & [ \% ] &  $\rocroc$ &  $\rocroc$ &  other $4\pi$  \\ \hline
1.10 -- 1.40  & 0.6 & 25 & $ 4.9 \pm 1.8  \pm 1.3  $  & $ 3.9 \pm 1.5 \pm 1.1 $ & $ \phantom{0}9.0 \pm 2.4 \pm 1.7 $ \\ \hline
1.40 -- 1.65  & 0.9 & 18 & $ 6.7 \pm 1.6  \pm 1.3  $  & $ 7.6 \pm 1.9 \pm 1.5 $ & $ 14.8 \pm 2.7 \pm 2.5 $ \\ \hline
1.65 -- 1.85  & 1.1 & 15 & $ 5.1 \pm 1.5  \pm 0.9  $  & $ 8.4 \pm 2.4 \pm 1.6 $ & $ 15.8 \pm 3.1 \pm 2.3 $ \\ \hline
1.85 -- 2.10  & 1.1 & 13 & $ 3.9 \pm 1.4  \pm 0.8  $  & $ 5.9 \pm 2.0 \pm 1.2 $ & $ 18.3 \pm 3.0 \pm 2.7 $ \\ \hline
2.10 -- 2.40  & 1.2 & 10 & $ 2.2 \pm 1.0  \pm 0.5  $  & $ 3.2 \pm 1.4 \pm 0.8 $ & $ 11.5 \pm 2.1 \pm 1.8 $ \\ \hline
2.40 -- 3.00  & 1.2 & 10 & $ 2.2 \pm 1.0  \pm 0.5  $  & $ 1.9 \pm 0.9 \pm 0.5 $ & $ \phantom{0}8.5 \pm 1.5 \pm 1.5 $ \\ \hline
\end{tabular}
\caption{Detection efficiencies, $\varepsilon$,  background fractions, $\mathit{Bg}$,
         and cross sections 
         of the reactions $ \EE \to \EE \rocroc$, $\gamgam \to \rocroc$ 
	        and  of the sum of the
         rest of the contributing processes, other $4\pi$,
         as a function of  $\mgg$ for $0.2 \GeV^2 \le \q \le 0.85 \GeV^2$.
         The first uncertainties are statistical, the second systematic.
        An overall normalization uncertainty of $5 \% $ for the trigger is not included. }
\label{tbl:xsectwgg}
\end{center}
\end{table*}

%

%
\clearpage
\vfil


\clearpage
\vfil


\clearpage
\hskip -0.5cm
\begin{minipage}{.5cm}
\begin{center}
\large
\rotatebox{90}{Entries / 80 MeV}\\
\normalsize
\end{center}
\end{minipage}\begin{minipage}{16.5cm}

\vskip -1.0cm

{\epsfig{file=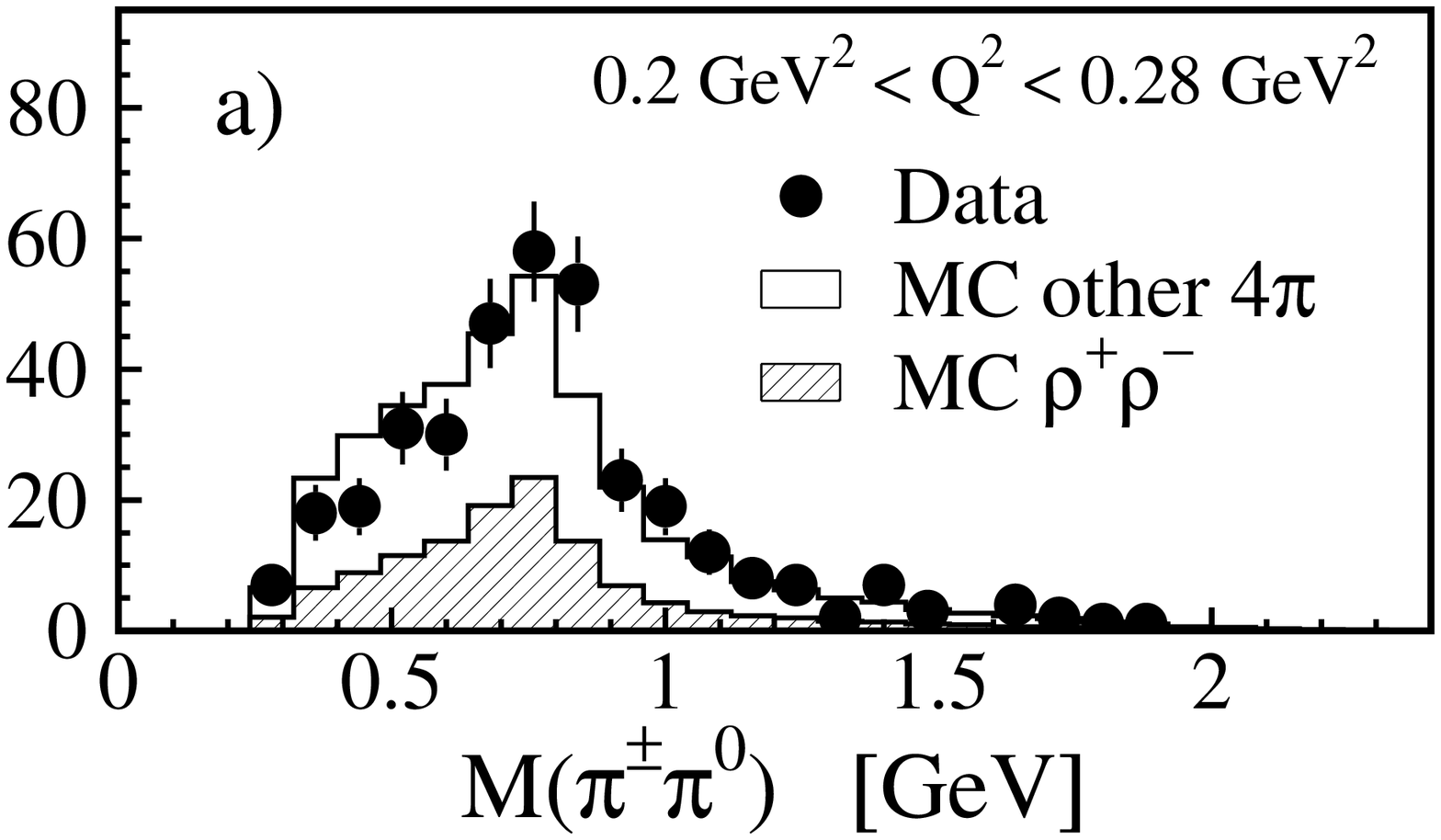,width=0.49\linewidth}}
{\epsfig{file=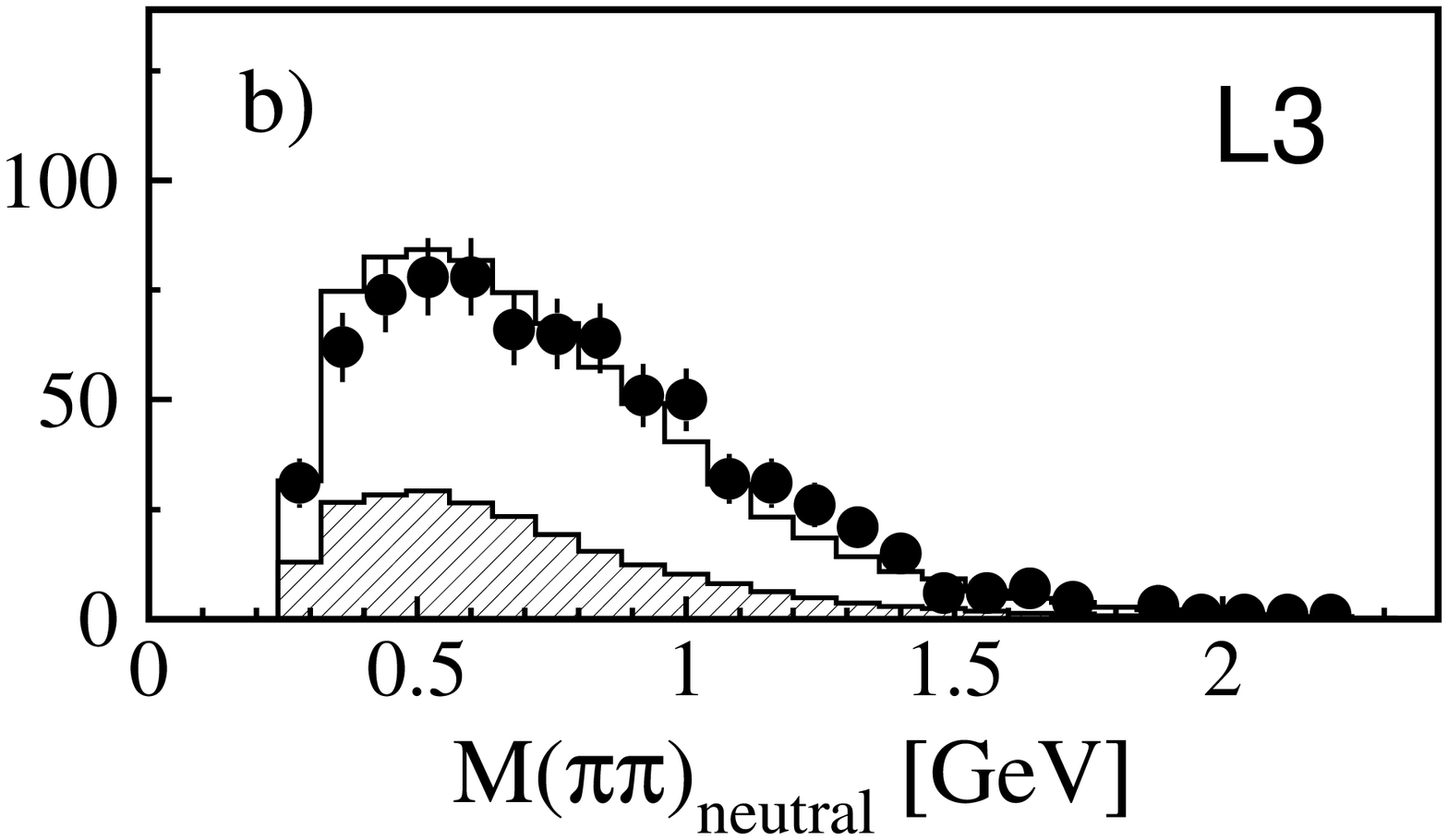,width=0.49\linewidth}}
\vskip -0.5cm
{\epsfig{file=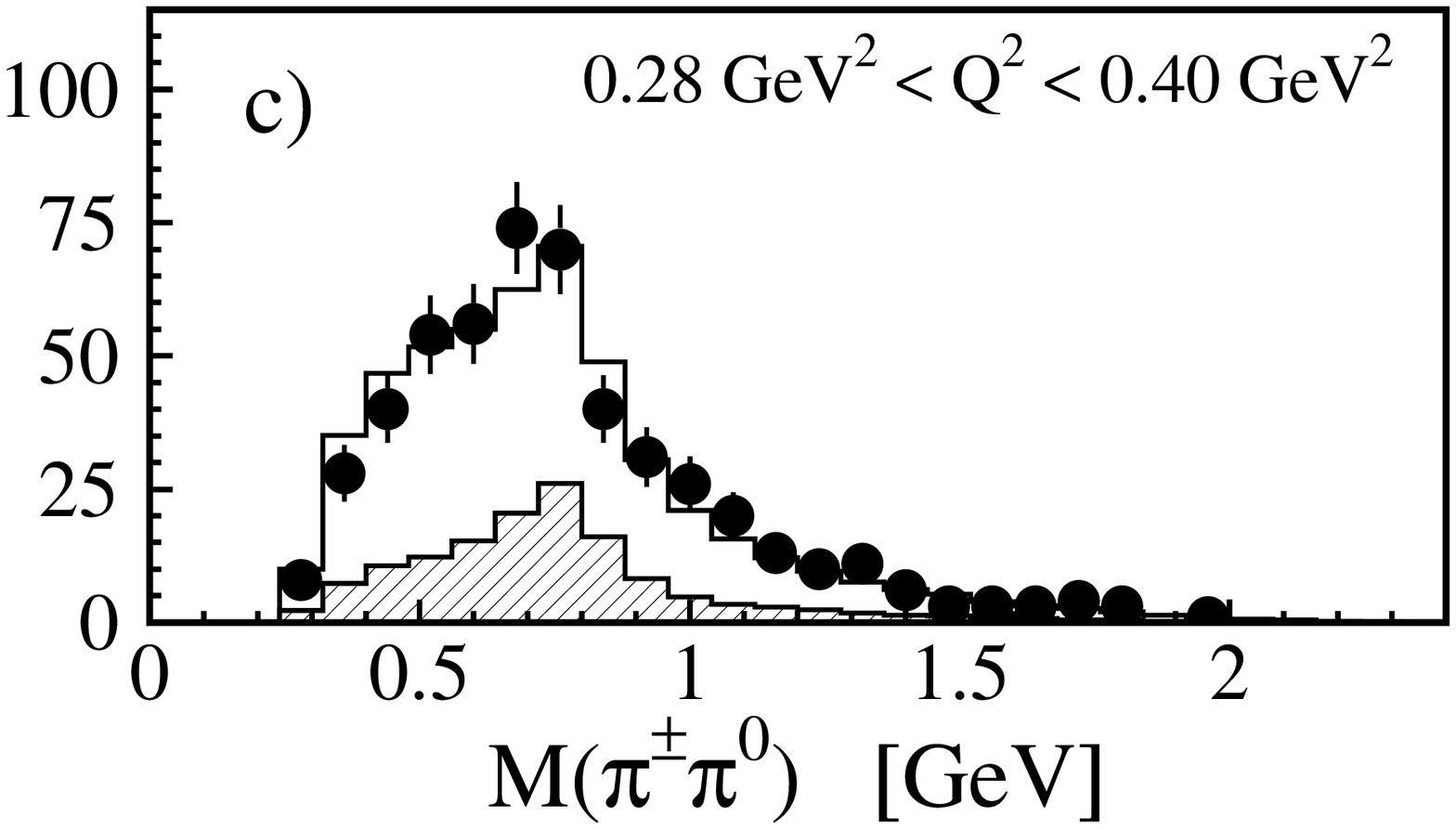,width=0.49\linewidth}}
{\epsfig{file=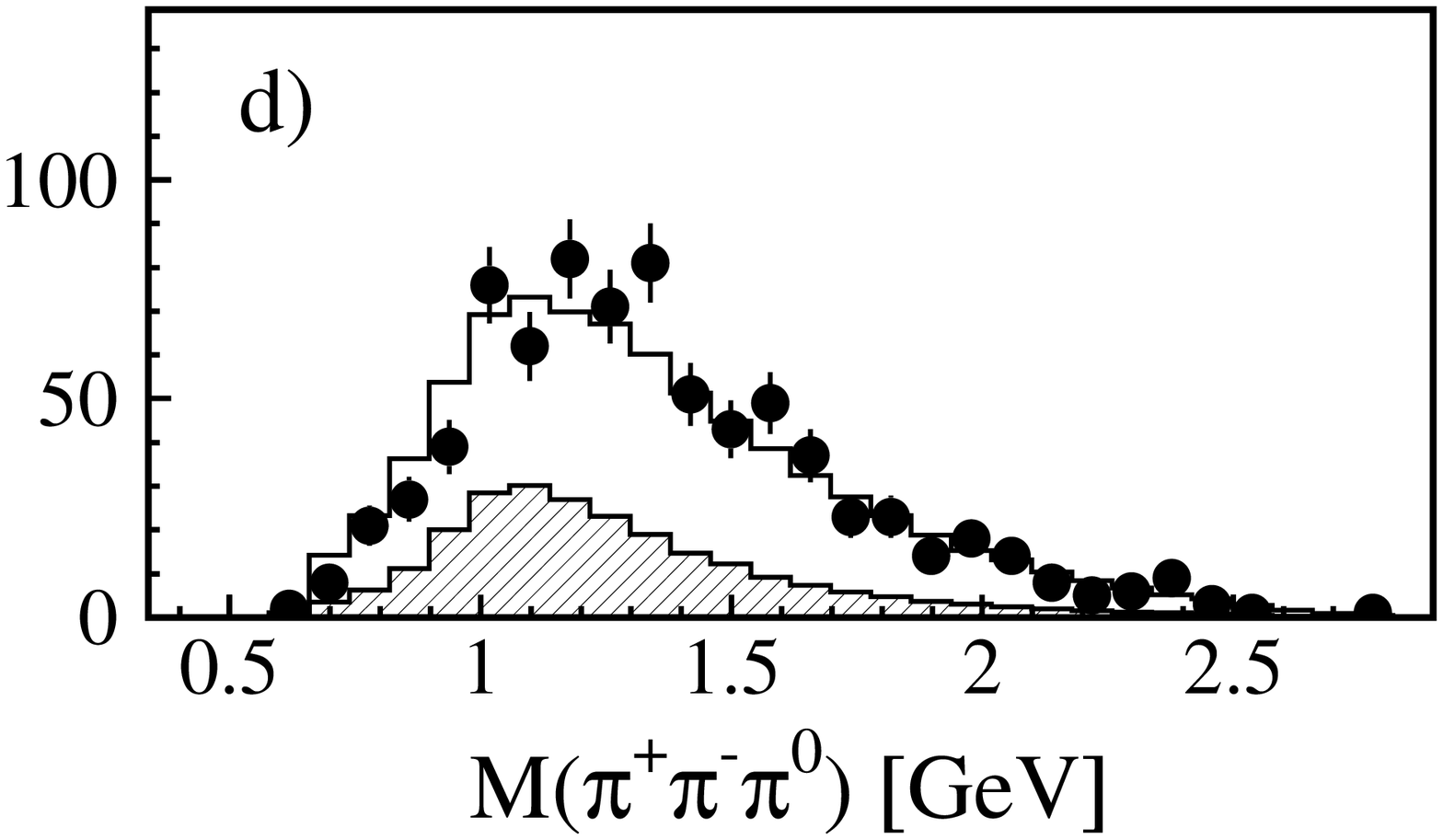,width=0.49\linewidth}}
\vskip -0.5cm
{\epsfig{file=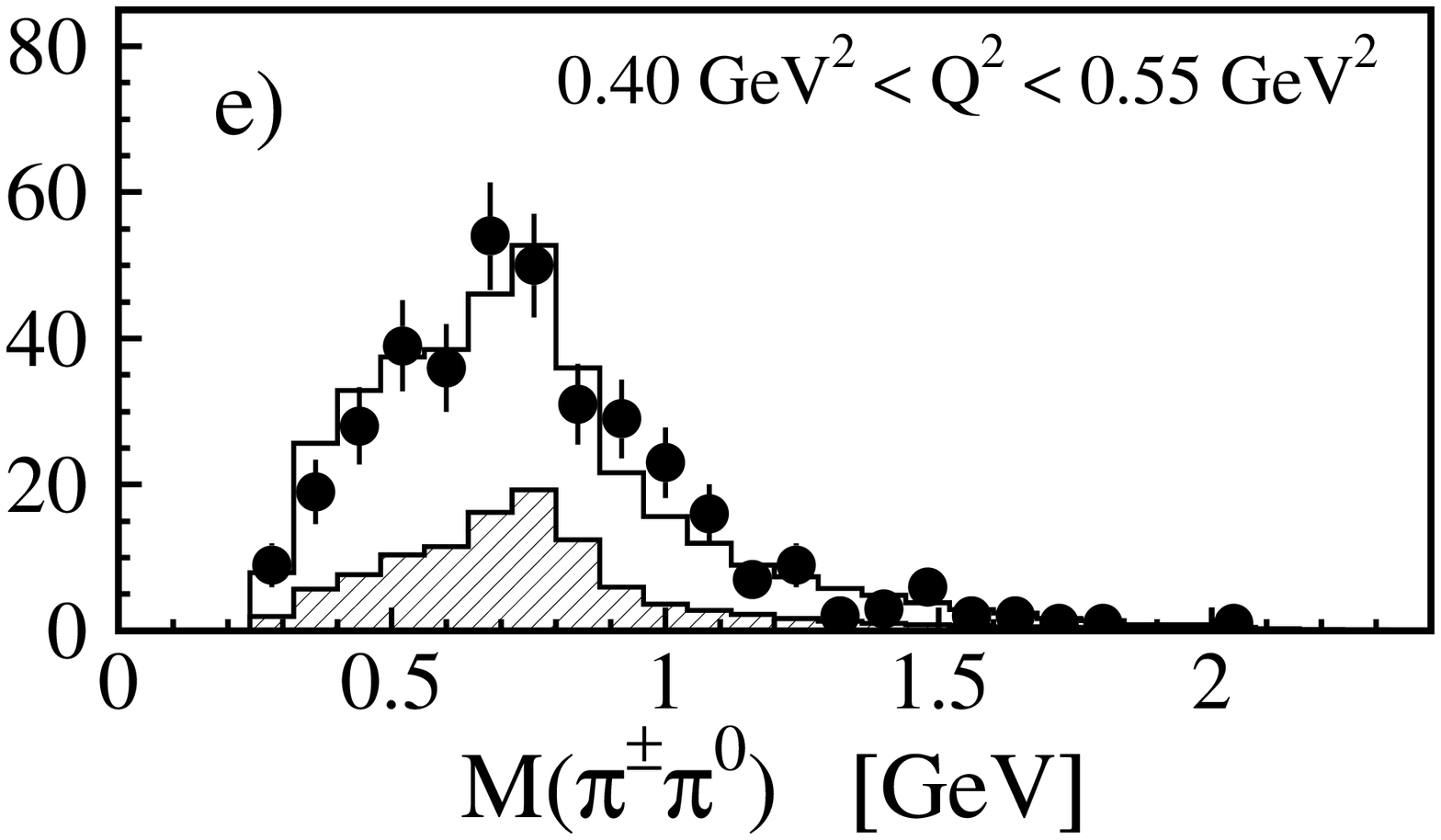,width=0.49\linewidth}}
{\epsfig{file=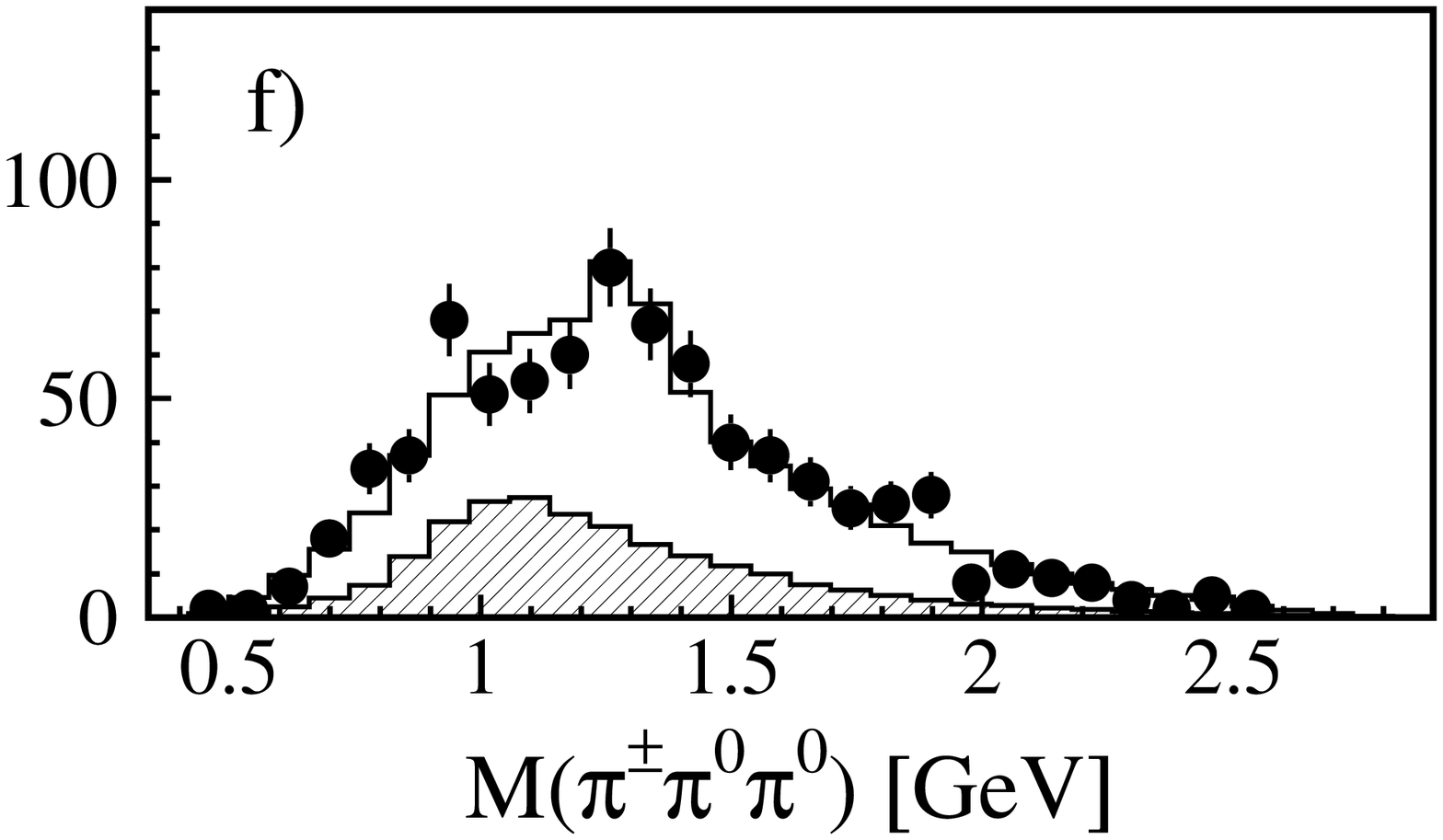,width=0.49\linewidth}}
\vskip -0.5cm
{\epsfig{file=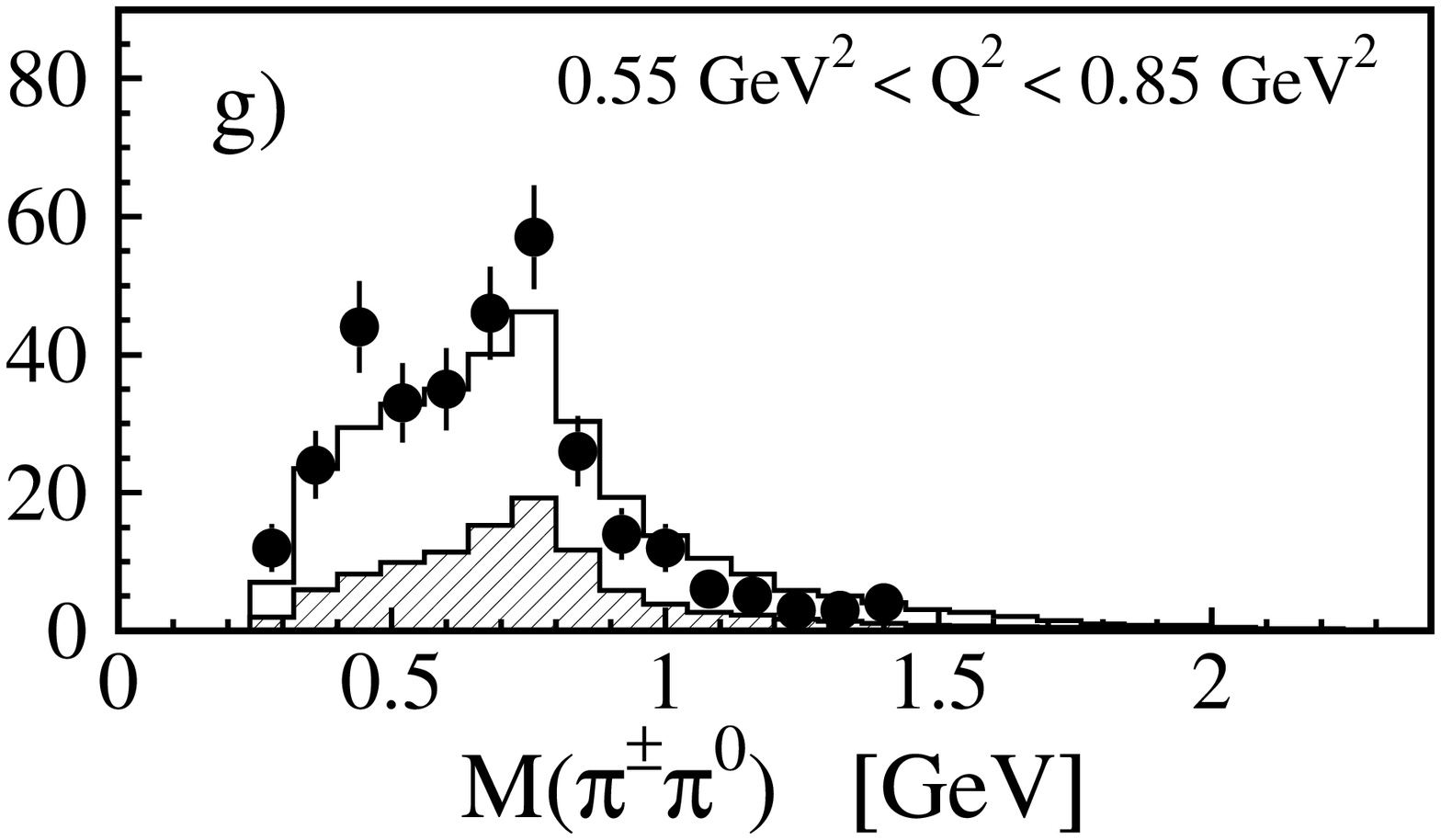,width=0.49\linewidth}}
{\epsfig{file=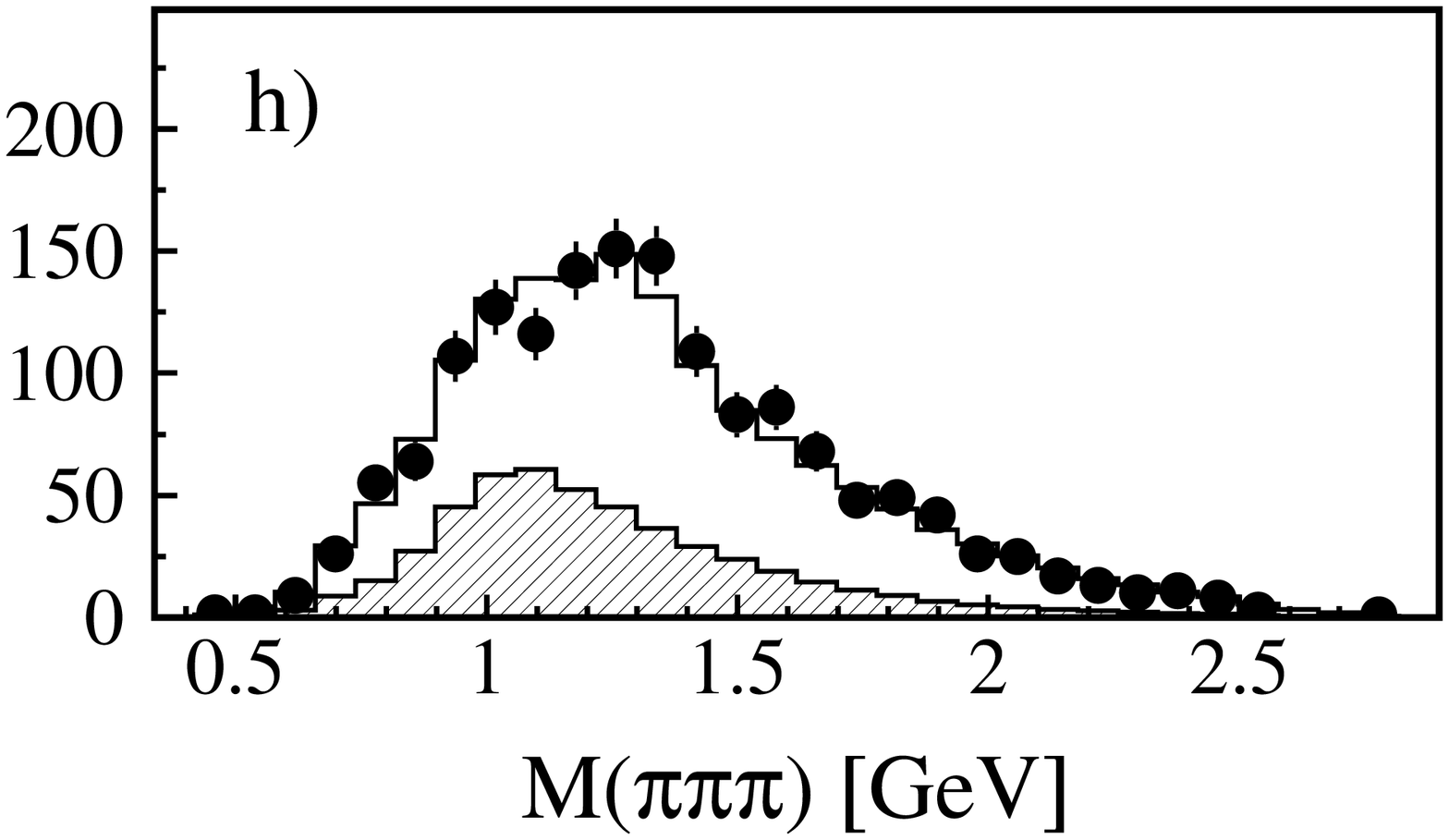,width=0.49\linewidth}}

\end{minipage}

\begin{figure}[h]
 \caption[]{ a),c),e),g) Mass distributions of the $\pi^\pm \piz$ combinations 
           (four entries per event) in  four $\q$-intervals and
           distributions for the entire kinematic region $1.1 \GeV  \le   \mgg \le  3 \GeV $
	    and  $0.2 \GeV^2  \le  \q  \le 0.85\GeV^2$ of
           b) The sum of the $\pipi$ and $\pizpiz$ mass spectra (two entries per event).
           d) The neutral three-pion combinations (two entries per event).
           f) The charged three-pion combinations (two entries per event).
           h) The sum of the $\pipi\piz$ and $\pi^\pm\pizpiz$ mass spectra (four entries per event).
                The points represent the data, the hatched areas show  the $\rocroc$ component
           and the open areas show the sum  of the other contributing processes.
           The fraction of the different components are determined by the fit and the
           total normalisation is to the number of the events.
       }
\label{fig:fig3}
\end{figure}
\vfil

 \begin{figure} [p]
  \begin{center}
    \mbox{\epsfig{file=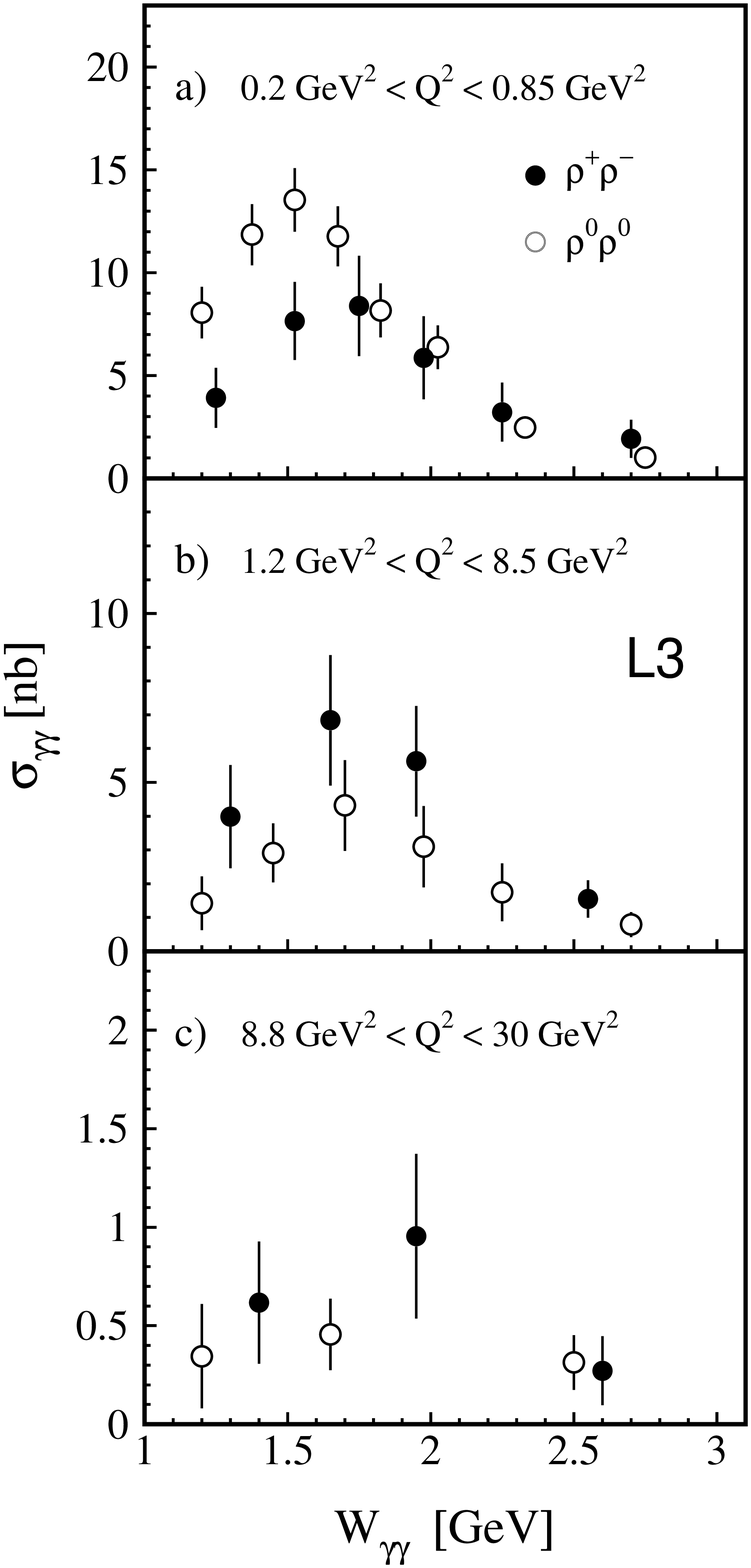,width=0.60\textwidth}}
  \end{center} 
  \caption[]{
            Cross section of the process $\gamgam\to\rho\rho$
            as  a function of  $\mgg$ in three $\q$ regions.  
            The results from this measurement,  full points in a), 
	    are compared to
             previous L3 measurements
            of the $\rho\rho$ production \protect\cite{L3paper269,L3paper287,L3paper292}.
            The  bars show the statistical uncertainties.
            Some points from the previous measurements are displaced horizontally for better readability.
           }
\label{fig:fig5}
\end{figure}
%
%
%

\clearpage
  \begin{figure} [p]
  \begin{center}
    \mbox{\epsfig{file=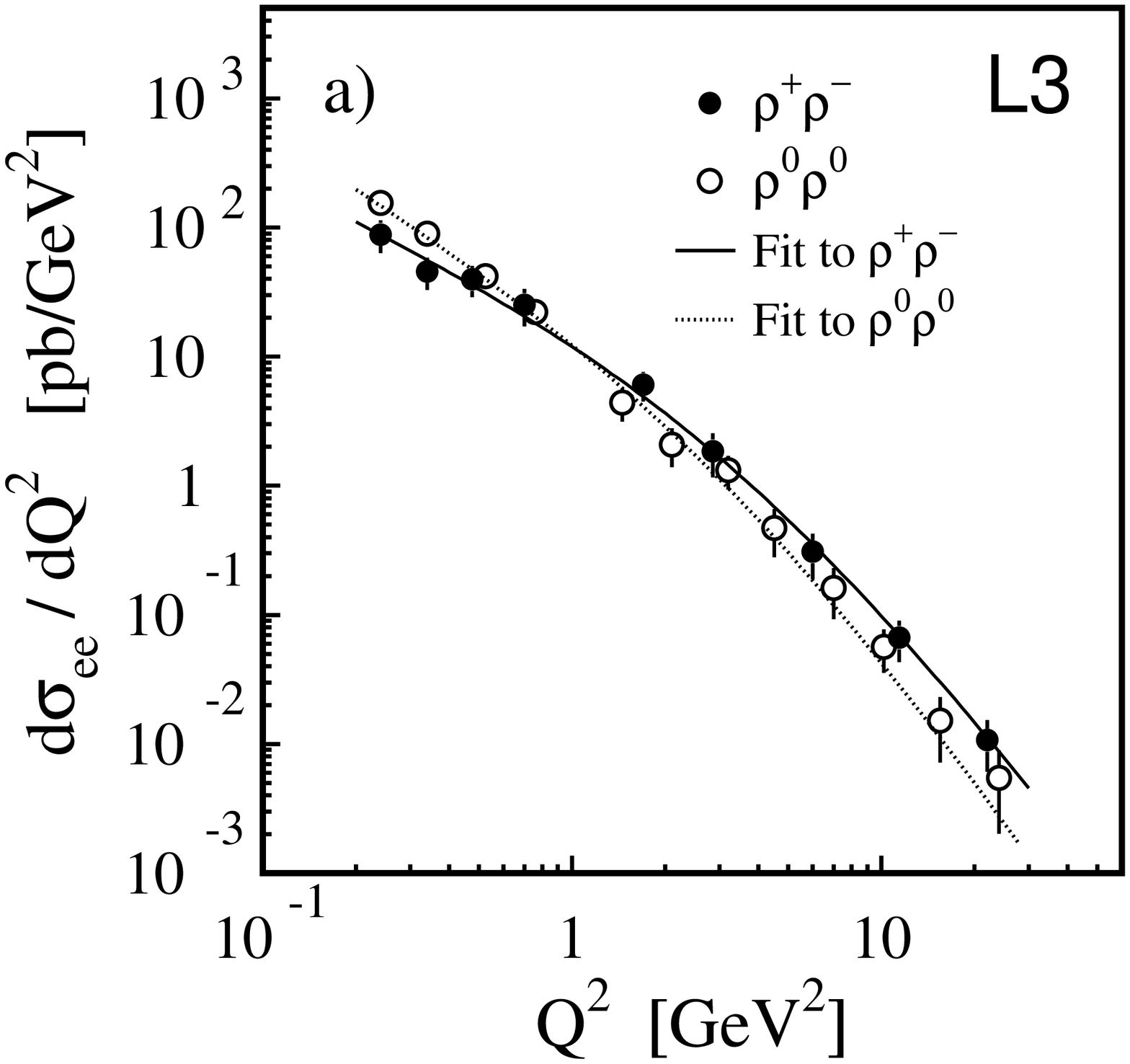,width=0.49\textwidth}}
    \mbox{\epsfig{file=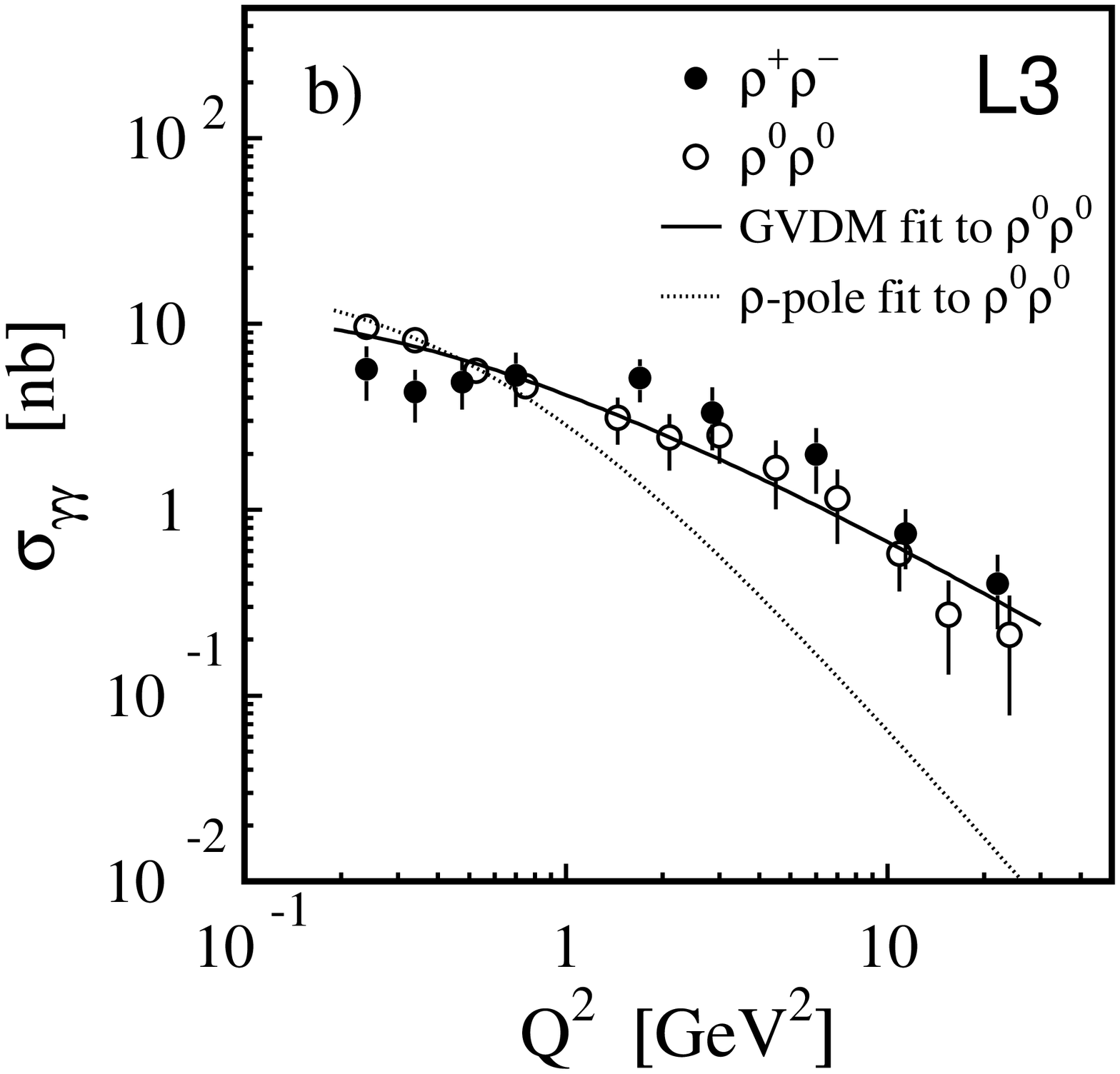,width=0.49\textwidth}}
  \end{center}
 \caption{The $\rho\rho$ production cross section as a function of $\q$,
         for $1.1 \GeV \le \mgg \le 3 \GeV$:
         a) differential cross section of the process
         $ \EE \to \EE \rho\rho$ and
         b) cross section of the process $\gamgam \to \rho\rho$.
          The results from this measurement,
	 full points in the region $\q < 1 \GeV^2$, are presented together with
          previous L3 measurements
         of the $\rho\rho$ production
         \protect\cite{L3paper269,L3paper287,L3paper292}.  The bars
         indicate the statistical uncertainties.  Some points from the
         previous measurements are displaced horizontally for better
         readability.  The lines in a) represent the results of fits
         using the QCD-inspired form of equation (9).  The lines in b)
         represent the results of a fit to the $\roro$ data based on
         the GVDM model \protect\cite{GVDM} and of a fit based on a
         $\rho$-pole parametrisation.
           }
\label{fig:fig6}
\end{figure}
\vfil

%
\end{document}